\documentclass[aps,prd,reprint,nofootinbib,showpacs]{revtex4-1}
\usepackage{amssymb}
\usepackage{amsmath}
\usepackage{amsfonts}
\usepackage{hyperref}
\usepackage{graphicx}
\usepackage{gensymb}
\usepackage{xcolor}

\begin{document}
\title{Testing nonlinear vacuum electrodynamics with Michelson interferometry}
\author{Gerold O. Schellstede}
\email{gerold.schellstede@zarm.uni-bremen.de \\ Alternative Email: schellst@physik.fu-berlin.de}
\affiliation{ZARM, University of Bremen, Am Fallturm, 28359 Bremen, Germany}
\author{Volker Perlick}
\email{volker.perlick@zarm.uni-bremen.de}
\affiliation{ZARM, University of Bremen, Am Fallturm, 28359 Bremen, Germany}
\author{Claus L\"{a}mmerzahl}
\email{claus.laemmerzahl@zarm.uni-bremen.de}
\affiliation{ZARM, University of Bremen, Am Fallturm, 28359 Bremen, Germany}
\affiliation{Institute of Physics, University of Oldenburg, 26111 Oldenburg, Germany}
%\date{}
\pacs{03.50.Kk,11.10.Lm}

%---------------------------------------------------------------------
\begin{abstract}
We discuss the theoretical foundations for testing nonlinear vacuum electro\-dynamics with Michelson 
interferometry. Apart from some nondegeneracy conditions to be imposed, our discussion applies to all 
nonlinear electrodynamical theories of the Pleba{\'n}ski class, i.e., to all Lagrangians that depend 
only on the two Lorentz-invariant scalars quadratic in the field strength. The main idea of the experiment 
proposed here is to use the fact that, according to nonlinear electrodynamics, the phase velocity of 
light should depend on the strength and on the direction of an electromagnetic background field. There 
are two possible experimental setups for testing this prediction with Michelson interferometry. The 
first possibility is to apply a strong electromagnetic field to the beam in one arm of the interferometer 
and to compare the situation where the field is switched on with the situation where it is switched off. 
The second possibility is to place the whole interferometer in a strong electromagnetic field and to 
rotate it. If an electromagnetic field is placed in one arm, the interferometer could have the size of 
a gravitational wave detector, i.e., an arm
length of several hundred meters. If the whole interferometer 
is placed in an electromagnetic field, one would have to do the experiment with a tabletop interferometer.
As an alternative to a traditional Michelson interferometer, one could use a pair of optical resonators 
that are not bigger than a few centimeters. Then the whole apparatus would be placed in the background 
field and one would either compare the situation where the field is switched on with the situation where
it is switched off or one would rotate the apparatus with the field kept switched on. We derive the 
theoretical foundations for these types of experiments, in the context of an unspecified nonlinear 
electrodynamics of the Pleba{\' n}ski class, and we discuss their feasibility. A null result of the 
experiment would place bounds on the parameters of the theory. We specify the general results to some 
particular theories of the Pleba{\' n}ski class; in particular, we give numerical estimates for Born, 
Born-Infeld and Heisenberg-Euler theories.
\end{abstract}
%---------------------------------------------------------------------------------------------------

\maketitle

\section{Introduction}\label{sec:intro}

Modifying an earlier idea by Born~\cite{Born1933}, in 1934 Born and Infeld \cite{BornInfeld1934} 
suggested a nonlinear modification of vacuum electrodynamics in order to get rid of the infinite 
self-energies of point particles that occur in the standard Maxwell theory. Their theory can be 
derived from a Lorentz-invariant Lagrangian. A few years later, Heisenberg and Euler~\cite{HeisenbergEuler1936} 
derived an effective Lagrangian, again Lorentz-invariant, from quantum electrodynamics. These are the 
two best known examples within the class of all Lorentz-invariant nonlinear electrodynamical 
theories. More generally, Pleba{\'n}ski \cite{Plebanski1970} and also Boillat~\cite{Boillat} studied the whole class of nonlinear electrodynamical theories that can be derived from a Lagrangian depending only
on the two Lorentz-invariant scalars that are quadratic in the field strength. This class
is often referred to as \emph{Pleba{\'n}ski nonlinear electrodynamics}. 
For a review of the Born-Infeld theory we refer, e.g.,  to 
Bia{\l}ynicki-Birula~\cite{Bialynicki}.

The physical relevance of these nonlinear vacuum electrodynamical theories is being widely 
discussed in the literature. It is believed that at a certain field strength the
Heisenberg-Euler deviations from standard Maxwell theory should be observable, and the 
Born-Infeld theory has gained increasing attention since it was realized by Tseytlin \cite{Tseytlin1999} 
that the Born-Infeld Lagrangian can be derived as an effective Lagrangian from some versions of 
string theory. 
Observable effects of (nonlinear) modifications of the 
vacuum Maxwell equations have been discussed for many years, at least
since the Ph.D. thesis of Toll~\cite{Toll:1952rq}.
Up to now, the only effect predicted by such modified theories
that has already been 
observed is light-by-light scattering (see \cite{Burkeetal1997});
further experiments are under way, e.g. with the Large Hadron Collider at CERN \cite{EnterriaSilveira2013}. 
There is also an ongoing experiment \cite{Dellavalleetal2013} aiming at verifying the birefringence 
\textit{in vacuo} as predicted by the Heisenberg-Euler theory. Also, it might be possible to measure the 
influence of background fields on the propagation speed of light in the laboratory. For the case 
of the Born-Infeld theory, such experiments have been suggested with the help of wave guides by 
Ferraro \cite{Ferraro2007} and with homogeneous magnetic background 
fields by Dereli and Tucker \cite{DereliTucker2010}. 

In this paper we focus on another method for testing nonlinear electrodynamics and discuss its theoretical 
foundations in detail. The basic idea is to measure the influence of a (strong) background field on the 
propagation speed of light with the help of an interferometer. 
Such an experiment has been suggested and discussed
already in five earlier papers~\cite{Boer,Denisov,Doebrich,Zavattini,Grote}.
However, all of them restrict the theoretical discussion to the 
Heisenberg-Euler theory or, in the case of Denisov \textit{et al.} \cite{Denisov},
to the Heisenberg-Euler and the Born-Infeld theory.
What is still missing is a comprehensive derivation of the relevant 
equations that cover the whole Pleba{\'n}ski class.
The basic idea of the experiment is simple. In the standard Maxwell vacuum theory, which is linear, the superposition principle holds, so 
there is no influence of a background field on the propagation of light. In the nonlinear theories, however, 
the phase velocity of light depends on the strength of the background field and on the propagation direction 
relative to the background field. This can be tested with a Michelson interferometer: If a strong background 
field is switched on and off in one interferometer arm, or if the whole interferometer is being rotated in a 
strong background field, the interference pattern should change. A null result would place bounds on the possible 
deviations from standard Maxwell vacuum theory. 

It is the purpose of this paper to develop the theoretical foundations for this experimental test for an 
unspecified nonlinear electrodynamical theory of the Pleba{\'n}ski class. We will then specify to Born,
Born-Infeld and Heisenberg-Euler theory. 

Throughout this paper, we consider Minkowski space as the underlying space-time model. We work 
in inertial coordinates, so the Minkowski metric is $(\eta ^{ik})=\mathrm{diag}(1,1,1,-1)$. We 
use Einstein's summation convention for Latin indices taking values 1,2,3,4 and for Greek 
indices taking values 1,2,3. Indices are raised and lowered with the Minkowski metric. We will use 
Gauss\-ian cgs units throughout, because they are most convenient for our theoretical investigations.
In these units, $E$, $B$, $D$ and $H$ are all measured in the same units, 
$\sqrt{\mathrm{g}}/ \big(\sqrt{\mathrm{cm}} \, \mathrm{s} \big)$.
The reader can easily convert the results into SI units with the help of the formulas 
$E = \sqrt{4 \pi \epsilon _0} E_{\mathrm{SI}}$, 
$B = \sqrt{4 \pi / \mu _0} B_{\mathrm{SI}}$, 
$D = \sqrt{4 \pi / \epsilon _0} D_{\mathrm{SI}}$ and  
$H = \sqrt{4 \pi \mu _0} H_{\mathrm{SI}}$ \cite{Jackson5te}.
For example, for a field $X = 10^3 \sqrt{\mathrm{g}} / \big( \sqrt{\mathrm{cm}} \, \mathrm{s} \big)$
in Gaussian cgs units, where $X = E$, $B$, $D$, or $H$, one gets
\begin{equation}\label{eq:SI}
\begin{split}
E_{\mathrm{SI}} = 3 \times 10^7 \, \dfrac{\mathrm{V}}{\mathrm{m}} 
\, , \qquad
B_{\mathrm{SI}} = 100 \, \mathrm{mT}
\, , \quad
\\
D_{\mathrm{SI}} = 3 \times 10^{-4} \, \dfrac{\mathrm{As}}{\mathrm{m}^2}
\, , \qquad
H_{\mathrm{SI}} = 8 \times 10^4 \, \dfrac{\mathrm{A}}{\mathrm{m}}
\, .
\end{split}
\end{equation}

The paper is organized as follows. In Sec.~\ref{Vorbereitung} we recall the basic equations 
for the propagation of light rays according to nonlinear electrodynamics. Then in 
Sec.~\ref{experiment} the suggested interferometer experiment is described and in 
Sec.~\ref{konkretes} some particular applications are discussed.

%-------------------------------------------------------------------------------------------------------------------------------------------------------
\section{Light propagation in nonlinear electrodynamics}\label{Vorbereitung}

\subsection{The Pleba{\'n}ski class of nonlinear electrodynamical theories}\label{subsec:Pleb}
The nonlinear electrodynamical theories which are at the center of our examination derive from an action
%\footnote{For simplicity the Lagrangian of the free electromagnetic field is named $L$. 
%The Lagrangian of the free field plus the coupling term is named $L_\text{C}$.}
\begin{equation}
S[A_m]=\frac{1}{4\pi c}\int _M \left(\mathcal{L}(F_{mn})+\frac{4\pi}{c}\,j^mA_m \right)\,\mathrm dV_4 \,.
\end{equation}
Here $j^m$ is a \emph{given} current density, $A_m$ is the electromagnetic potential, 
$F_{mn} = \partial _m A_n-\partial _n A_m$ is the electromagnetic field strength and 
$\mathcal{L}$ is the Larangian for the electromagnetic field. Then the homogeneous 
group of Maxwell's equations is automatically satisfied,
\begin{equation}\label{eq:Max1}
\partial_{[a} F_{bc]}=0 \, .
\end{equation}
Variation of the action with respect to the potential $A_m$ leads to the inhomogeneous 
group of Maxwell's equations,
\begin{equation}\label{eq:Max2}
\partial_b H^{ab} \, = \, \frac{4\pi}{c}\,j^a\,,
\end{equation}
where
%\footnote{Sometimes we make use of the following short notations: 
%$L_F:=\frac{\partial L}{\partial F}$, 
%$L_{FG}:=\frac{\partial^2}{\partial F\partial G}$ and so on.}  
\begin{equation}\label{eq:const}
H^{ab}=-\frac{\partial \mathcal{L}}{\partial F_{ab}}
\end{equation}
is the electromagnetic excitation. It is the constitutive law (\ref{eq:const}) that 
distinguishes  different theories, while the Maxwell equations 
(\ref{eq:Max1}) and (\ref{eq:Max2}) are always the same.
 
Following Pleba{\'n}ski \cite{Plebanski1970}, we require that the electromagnetic 
Lagrangian $\mathcal{L}$ depends on the electromagnetic field strength only via the Lorentz 
invariants
\begin{equation}\label{eq:FG}
F=\frac{1}{2}\,F_{mn}F^{mn} 
\quad \text{and} \quad
G=-\frac{1}{4}\,F_{mn}\tilde F^{mn} \, .
\end{equation}
Here and in the following, the tilde denotes the Hodge dual,
\begin{equation}
\tilde F^{mn}\, = \, \frac{1}{2} \, \varepsilon^{mnab}F_{ab} \, .
\end{equation}
%\footnote{All following duals are denoted by $\sim$.} 
As usual, $\varepsilon^{abcd}$ is the totally antisymmetric Levi-Civita 
tensor with $\varepsilon^{1234}=-1$. Strictly speaking, only $F$ is 
invariant under \emph{all} Lorentz transformations while $G$ changes
sign under a parity transformation. Some authors restrict to Lagrangians 
that satisfy the equation $\mathcal{L}(F,G)=\mathcal{L}(F,-G)$ to assure 
invariance under parity transformations. However, for the purpose of this 
paper there is no need for this restriction.

The Pleba{\'n}ski class contains, of course, the standard vacuum
Maxwell theory which is given by the Lagrangian
\begin{equation}\label{eq:LagMax}
\mathcal{L}{}_M (F,G) = \, - \, \dfrac{1}{2} \, F \, .
\end{equation}
As this theory is well tested for weak fields, many authors restrict their work to 
theories where the Lagrangian satisfies $\mathcal{L} + F/2 \to 0$ for
$F_{mn} \to 0$. Again, for our mathematical considerations there is no
need for this restriction.

For a theory of the Pleba{\'n}ski class the constitutive law
(\ref{eq:const}) can be written, more specifically, as 
\begin{equation}\label{eq:Hab}
H^{ab}= -2\, \mathcal{L}_F \, F^{ab}+\mathcal{L}_G \,\tilde F^{ab} 
\end{equation}
where
\begin{equation}\label{eq:LFLG}
\mathcal{L}_F = \dfrac{\partial \mathcal{L}}{\partial F} \, , \quad
\mathcal{L}_G = \dfrac{\partial \mathcal{L}}{\partial G} \, .
\end{equation}
Later we will also write
\begin{equation}\label{eq:LFLG2}
\mathcal{L}_{FF} = \dfrac{\partial ^2 \mathcal{L}}{\partial F ^2} \, , \quad
\mathcal{L}_{GG} = \dfrac{\partial ^2 \mathcal{L}}{\partial G ^2} \, , \quad
\mathcal{L}_{FG} = \dfrac{\partial ^2 \mathcal{L}}{\partial F \partial G} \, .
\end{equation}

Additionally we will use a Hamiltonian formulation of the Pleba{\' n}ski electrodynamics. 
For the special case of the Born-Infeld theory, 
the Hamiltonian formulation can be found in~\cite{BornInfeld1934,Bialynicki}
Whereas the Lagrangian depends on the field strength, the Hamiltonian depends on the
excitation. Quite generally, for any theory based on a Lagrangian $\mathcal{L}(F_{mn})$,
the passage to the Hamiltonian formalism can be performed whenever the constitutive 
law (\ref{eq:const}) can be solved for $F_{mn}$. The Hamiltonian is then given by
a covariant Legendre transformation,
\begin{equation}\label{eq:Hamilton}
\mathcal{H}(H^{ab})=-\frac{1}{2}H^{mn}F_{mn}-\mathcal{L}(F_{ab})
\end{equation}
where, on the right-hand side, $F_{ab}$ has to be expressed in 
terms of $H^{mn}$ with the help of the constitutive law.
For a theory of the Pleba{\'n}ski class (i.e., if the Lagrangian
depends only on $F$ and $G$), the Hamiltonian is a function of
the two invariants 
\begin{equation}\label{eq:RS}
R=-\frac{1}{2} \,H^{ab}H_{ab}
\quad \text{and} \quad
S=\frac{1}{4}H_{ab}\tilde H{}^{ab} \, .
\end{equation}
The relevant equations for the passage from the Lagrangian to the
Hamiltonian description are given in the Appendix. There we will 
also give a criterion that guarantees that the constitutive law  
(\ref{eq:Hab}) can be solved for the field strength, at least 
locally.

%------------------------------------------------------------------------------------------------
\subsection{Three-dimensional notation of field equations}
In the following we will often use three-vector notation. The field strength has the  
three-dimensional representation
\begin{equation}\label{eq:EB}
E_{\alpha} = F_{\alpha 4}
\, , \quad
B^\alpha=\frac12\,\varepsilon^{\alpha\beta\gamma}F_{\beta\gamma} 
\, = \, - \, \tilde{F}{}^{\alpha 4} \, ,
%(F_{mn})=\left(\begin{array}{c|c}
%B_{\alpha\beta}&E_{\alpha}\\
%\hline 
%-E_{\beta}&0
%\end{array}\right)\quad\text{with}\quad B^\alpha=\frac12\,\varepsilon^{\alpha\beta\gamma}B_{\beta\gamma}\quad\Rightarrow\quad B^1=B_{23}\,,\,B^2=B_{31}\,,\,B^3=B_{12}\,.
\end{equation}
where $\varepsilon ^{\alpha \beta \gamma}$ is the totally antisymmetric 
spatial Levi-Civita tensor, $\varepsilon ^{123}=1$.
%For the four-dimensional current-density we write
%\begin{equation}
%(j^m)=(j^\alpha,c\rho)\,
%\end{equation}
%and for the four-potential
%\begin{equation}
%(A_a)=\left(A_\alpha,-\phi\right)\,.
%\end{equation}
With $E^2 = \delta ^{\mu \nu} E_{\mu}E_{\nu}$, 
$B^2 = \delta _{\mu \nu} B^{\mu}B^{\nu}$ and
$\mathbf{E} \cdot \mathbf{B} = E_{\mu}B^{\mu}$, the
invariants (\ref{eq:FG}) read
\begin{equation}\label{eq:EB1}
F=
%\frac{1}{2}
%\,B_{mn}B^{mn}=
B ^2- E ^2
\, , \quad
%&\text{and}&
G=
%-\frac{1}{4}\,B_{mn}\tilde B^{mn}=
\mathbf B \cdot \mathbf E \, .
\end{equation}
%\begin{equation}\label{eq:EB2}
%\mathbf E=-\text{grad}\,\phi-\frac{1}{c}\,\frac{\partial\mathbf A}{\partial t}
%\, , \quad
%&\text{and}& 
%\mathbf B=\text{curl}\,\mathbf A\,.
%\end{equation}
Analogously we write for the excitation
\begin{equation}\label{eq:DH}
D^{\alpha} = - H^{\alpha 4} \, , \quad
H_{\alpha}=\frac12\,\varepsilon_{\alpha\beta\gamma}H^{\beta\gamma} 
\, = \,
\tilde{H}{}_{\alpha 4} \, ,
%(H^{mn})=\left(\begin{array}{c|c}
%H^{\alpha\beta}&-D^{\alpha}\\
%\hline 
%D^{\beta}&0
%\end{array}\right)\quad\text{with}\quad 
%H_\alpha=\frac12\,\varepsilon_{\alpha\beta\gamma}H^{\beta\gamma}
%\quad\Rightarrow\quad H_1=H^{23}\,,\,H_2=H^{31}\,,\,H_3=H^{12}\,.
\end{equation}
which implies that the invariants (\ref{eq:RS}) are given by
\begin{equation}\label{eq:DH1}
R= D^2-H^2 \, , \quad S= \mathbf{D} \cdot \mathbf{H} \, .
\end{equation}
Then the constitutive law (\ref{eq:Hab}) reads
\begin{equation}
\begin{split}
&D_{\alpha}=-2 \mathcal{L} _F E_{\alpha} \,+ \mathcal{L}_G B_{\alpha}  \, ,\\
&H_{\alpha} =-2 \mathcal{L} _F B_{\alpha} \,- \mathcal{L}_G E_{\alpha} \, \,.
\end{split}
\end{equation}

%-----------------------------------------------------------------------------------------------------------------------------
\subsection{Phase velocity and characteristic differential equation for 
$\mathbf{\mathcal{L}(F,G)}$ theories}
The characteristic surfaces determined by a set of partial differential equations can be 
defined as the hypersurfaces along which the solutions may have discontinuities. As an 
alternative, the characteristic surfaces can also be defined with the help of approximate 
plane waves; in this second approach, they come about as the high-frequency limit of the 
surfaces of constant phase. In view of applications to electrodynamics, the first approach 
is discussed, e.g., in the book by Hehl and Obukhov \cite{HehlObukhov2003}. The characteristic 
surfaces are hypersurfaces $\psi = \mathrm{constant}$, where the gradient of $\psi$ has to 
satisfy, at each point of space-time, a fourth-order equation which is known as the 
\emph{dispersion relation} or as the  \emph{Fresnel equation}. If viewed as a partial 
differential equation for $\psi$, this equation is usually called the \emph{characteristic 
equation} or the \emph{eikonal equation}. Using this approach, Obukhov and 
Rubilar~\cite{ObukhovRubilar2002} have determined the Fresnel equation (i.e., the
characteristic equation) for an arbitrary $\mathcal{L}(F,G)$ theory. Earlier, 
Novello \textit{et al.}~\cite{Novello} had found an equivalent result in a different way. 
Their results show that, with the exception of a few special cases, theories 
of the Pleba{\'n}ski class predict birefringence \textit{in vacuo}. For background material on 
birefringence, and bimetricity, we refer to Visser \textit{et al.}~\cite{Visser} and, for the 
particular case of the Heisenberg-Euler theory, to Dittrich and Gies~\cite{Dittrich} 
and to Shore~\cite{Shore}.

Here we want to briefly sketch how the Fresnel equation of an arbitray $\mathcal{L} (F,G)$
theory can be derived with the help of an approximate-plane-wave ansatz. This is 
methodically different from the work of Obukhov and Rubilar~\cite{ObukhovRubilar2002} 
and Novello \textit{et al.}~\cite{Novello} but it leads to the same result.
The general method goes back to Luneburg and is outlined, 
e.g., for electrodynamics in ordinary media, in the book by Kline and Kay~\cite{KlineKay1965}. 
For a discussion in a more general context, which includes the case to be considered here, 
we refer to Perlick~\cite{Perlick2011}. 

We consider a one-parameter family of electromagnetic fields of the form
\begin{equation}\label{eq:pw}
\begin{split}
&F^{'ab} (x^m)=F^{ab} (x^m) \\
	&\quad\quad+ \mathrm{Re} \left\{
e^{-i\psi(x^m)/\lambda}
\sum\limits_{N=1}^\infty
\left(\lambda^N F_N^{ab} (x^m)\right)
\right\} \,,
\end{split}
\end{equation}
where $F^{ab}$ is a given background field. $\lambda$ is a real bookkeeping parameter
that is introduced in a way such that the high-frequency limit corresponds to 
$\lambda \to 0$. The summation sign in (\ref{eq:pw}) is to be understood in the 
sense of an asymptotic series and \emph{not} in the sense of a convergent series.
While the amplitudes $F_N^{ab}$ are in general complex, the \emph{eikonal function}
$\psi$ is real. It gives the 
surfaces of constant phase, $\psi(x^m)=\psi(x^\mu,t)=\text{constant}$. 
In 3-space, the normal to these surfaces is 
\begin{equation}
n_\alpha=\frac{\partial_\alpha\psi}{\sqrt{\left(\partial_\beta\psi\right)
\left(\partial^\beta\psi\right)}}\,.
\end{equation}
The phase velocity $v_{\text{P}}^{\alpha}$ can be introduced as the 3-vector 
that gives the traveling speed of such a surface in the direction 
of its normal,
\begin{equation}\label{eq:vP}
v_{\mathrm{P}}^{\alpha} = - \frac{\partial ^{\alpha} \psi }
{\left(\partial_\beta\psi\right)\left(\partial^\beta\psi\right)} \, 
\dfrac{\partial \psi}{\partial t}\,.
\end{equation}  
Feeding the ansatz (\ref{eq:pw}) into Maxwell's equations and comparing equal
powers of $\lambda$ gives a hierarchy of equations. In the lowest nontrivial 
order, which is known as the geometric optics approximation, one gets a 
first-order partial differential equation for $\psi$ which is the desired 
characteristic equation.   

If this program is carried through for an $\mathcal{L}(F,G)$ theory, one 
finds the following result which is in agreement with Obukhov and 
Rubilar's~\cite{ObukhovRubilar2002}. The characteristic equation reads
\begin{equation}\label{eq:Q}
\begin{split}
&\mathcal{L}_F \Big\{ M\eta^{ij}\eta^{kl}+N\eta^{ij}F^{km}F^{l}_{\phantom lm} 
\\
&\quad\qquad
+PF^{im}F^{j}_{\phantom jm}F^{kn}F^{l}_{\phantom ln} \Big\} p_ip_jp_kp_l=0
\end{split}
\end{equation}
where $p_i=\partial_i\psi$ and
%\begin{equation}\label{eq:sigma}
%Q_{1/2}=\left[\eta^{ij}+\left(\frac{N}{2M}\pm\sqrt{\frac{N^2}{4M^2}-\frac{P}{M}}\right)F^{im}F^{j}_{\phantom jm}\right]p_ip_j \, ,
%\quad p_i=\partial_i\psi\, ,
%\end{equation}
\begin{equation}
\begin{split}
&M=\mathcal{L}_F^2+2\,\mathcal{L}_F \mathcal{L}_{FG}\,G
-\frac12\,\mathcal{L}_F\mathcal{L}_{GG}\,F 
\\
&\quad\qquad+\left(\mathcal{L}^2_{FG}
-\mathcal{L}_{FF}\mathcal{L}_{GG}\right)G^2 \, ,
\\
&N=2\,\mathcal{L}_F \mathcal{L}_{FF}
+\frac12\,\mathcal{L}_F\mathcal{L}_{GG}
\\
&\quad\qquad+\left(\mathcal{L}^2_{FG}
-\mathcal{L}_{FF}\mathcal{L}_{GG}\right)F \, ,
\\
&P=\mathcal{L}_{FF}\mathcal{L}_{GG}-\mathcal{L}^2_{FG}\,.
\end{split}
\end{equation}
If $M$ has no zeros, (\ref{eq:Q}) can be factorized as 
%In the special case $M=0$ one gets another decomposition:
\begin{equation}\label{eq:a1a2}
\mathcal{L}_F \, M \, \big( a_1^{ij}p_ip_j \big) 
\big( a_2^{k \ell} p_k p_{\ell} \big) = 0
% Q=L_F\,Q_{1}\cdot Q_{2}=0\quad\text{where}\quad 
% Q_{1}=\left[N\eta^{ij}+P F^{im}F^{j}_{\phantom jm}\right]p_ip_j
% \quad\text{and}\quad Q_2=F^{im}F^{j}_{\phantom jm}p_ip_j\,,
\end{equation}
where 
\begin{equation}\label{eq:aA}
a_A^{ik} = \eta^{ik}+\sigma_A F^{im}F^k_{\phantom km}
\end{equation}
for $A=1,2$ and 
\begin{equation}\label{eq:sigmaA}
\sigma_{1/2} = \frac{N}{2M}\pm\sqrt{\frac{N^2}{4M^2}-\frac{P}{M}} \, .
\end{equation}
In the following we restrict ourselves to Lagrangians such that 
$M$ and $\mathcal{L}_F$ have no zeros. This excludes some degenerate
cases which are hardly of physical interest. Then the characteristic equation is
equivalent to (which is up to conformal transformations in agreement with the results of Novello \textit{et al.}\cite{Novello})
\begin{equation}\label{eq:cones}
a_A^{ik}p_ip_k=0 \, , \quad A=1,2 \, 
\end{equation}
and is sometimes called the "light-cone condition" (compare for 
example~\cite{Dittrich}). Generalizing a standard terminology from electrodynamics in media, 
$a_1^{ik}$ and $a_2^{ik}$ are called the \emph{optical metrics} of 
the vacuum in the $\mathcal{L}(F,G)$ theory. If the two optical metrics do
not coincide, i.e., if $\sigma _1 \neq \sigma _2$, there is birefringence
in vacuum. If one considers the next order in the above-mentioned hierarchy 
of equations, one sees that the case $A=1$ and the case $A=2$ correspond 
to two different polarization directions. 
Note that $\sigma_{1}$ and $\sigma_2$ are always real, because
\begin{gather}\label{eq:sigmareal}
N^2-4MP= 
\\
\nonumber
\Big( 2 \mathcal{L}_F \mathcal{L}_{FF} - 
\dfrac{1}{2} \mathcal{L}_F \mathcal{L}_{GG} - PF \Big) ^2
+ 4 \Big( \mathcal{L}_F \mathcal{L}_{FG}-PG \Big) ^2
\end{gather}
is a sum of two squares, and that $\sigma _1$ and $\sigma _2$ depend
only on the two field invariants $F$ and $G$. In the standard vacuum 
Maxwell theory we have $\sigma _1 = \sigma _2 =0$, so these
two functions characterize the deviation of our $\mathcal{L}(F,G)$ theory from 
the standard theory at the level of geometric optics.

%According to (\ref{eq:sigma}), $Q_1$ and $Q_2$ are quadratic forms in the $p_i$ which depend on the 
%electromagnetic field in a surprisingly simple way,
%\begin{equation}\label{eq:QA}
%Q_A=a_A^{ik}p_ip_k=\left(\eta^{ik}+\sigma_A F^{im}F^k_{\phantom km}\right)p_ip_k=0\, , \quad A=1,2 \,.
%\end{equation}

%Thereby one sees immediately that
%\begin{equation}
%N^2-4MP=\left(\left(L^2_{FG}-L_{FF}L_{GG}\right)F+L_F\left(2\,L_{FF}-\frac12\,L_{GG}\right)\right)^2+4\left(\left(L^2_{FG}-L_{FF}L_{GG}\right)G+L_FL_{FG}\right)^2
%\end{equation}
%secures that $\sigma_A$ is real function for all possible cases. 

%Now it is possible that the optical metric violates the causal structure. 
%In the Appendix \ref{KausalitaetOptMetr} we give a criterion to ensure that 
%this is not the case. In the former course we restrict ourselves to optical 
%metrics where such a violation does not occur. 

If the Lagrangian is of the special form $\mathcal{L}(F,G)=\mathcal{L}(\alpha F+\beta G)$ with
some constant factors $\alpha$ and $\beta$, one has $P= 0$ and therefore $\sigma_{1}=0$,
i.e., one  polarization mode behaves as in the standard Maxwell vacuum theory.
This is true, in particular, if the Lagrangian is independent of $G$. (It is 
also true if the Lagrangian is independent of $F$ but this case was excluded
by our assumption $\mathcal{L}_F \neq 0$.) Also, it is interesting to remark that 
two Lagrangians $\mathcal{L}$ and  $\mathcal{L}+\beta G$ give the same characteristic 
equation, i.e. the two cases are not distinguishable at the level of 
geometrical optics. Of course, if one restricts to parity invariant
Lagrangians adding a term of the form $\beta G$ is forbidden.

For some of our applications it will be desirable to write the optical
metrics in terms of the excitation, rather than in terms of the field
strength. It is then recommendable to start from a Hamiltonian formulation.
It was mentioned already at the end of Sec.~\ref{subsec:Pleb} that
the Pleba{\'n}ski class of theories can be written in terms of a 
Hamiltonian $\mathcal{H}(R,S)$ rather than in terms of a Lagrangian
$\mathcal{L}(F,G)$. In the Appendix we derive some replacement rules of
how the relevant Hamiltonian expressions can be found from the Lagrangian
expressions. By applying these replacement rules, we find that the 
optical metrics can be rewritten as
%\begin{equation}
%Q=Q_1\cdot Q_2=0\,
%\end{equation}
%where
\begin{equation}\label{eq:aAH}
 %Q_A=\hat 
 a{}_A^{ik}
 %p_ip_k
 =
 %\left(
 \eta^{ik}+\hat\sigma_A \tilde{H}{}^{im}
 \tilde{H}{}^k_{\phantom km}
 %\right)p_ip_k=0\, , \quad A=1,2 \,.
\end{equation}
where
\begin{equation}\label{eq:sigmaHam}
\hat{\sigma}{}_A=
\frac{\hat N}{2\hat M}\pm\sqrt{\frac{\hat N^2}{4\hat M^2}-\frac{\hat P}{\hat M}}
% Q_{1/2}=\left[\eta^{ij}+\left(\frac{\hat N}{2\hat M}\pm
%\sqrt{\frac{\hat N^2}{4\hat M^2}-\frac{\hat P}{\hat M}}\right)
%\hat H{}^{im}\hat H{}^{j}_{\phantom jm}\right]p_ip_j \, ,
%\quad p_i=\partial_i\psi\, ,
\end{equation}
with the abbreviations 
%\begin{equation}
% Q_A=\hat a{}_A^{ik}p_ip_k=\left(\eta^{ik}+\hat\sigma_A \tilde H{}^{im}\tilde H{}^k_{\phantom km}\right)p_ip_k=0\, , \quad A=1,2 \,.
%\end{equation}
%and
\begin{equation}
\begin{split}
&\hat M=\mathcal{H}_R^2+2\, \mathcal{H} _R \mathcal{H}_{RS}
\,S-\frac12\,\mathcal{H}_R 
\mathcal{H}_{SS}\,R
\\ & \quad\qquad+
\left(\mathcal{H}^2_{RS}-\mathcal{H}_{RR} \mathcal{H}_{SS}\right)S^2 \, ,
\\
&\hat N=2\,\mathcal{H}_R \mathcal{H}_{RR}
+\frac12\,\mathcal{H}_R \mathcal{H}_{SS}
\\ &\quad\qquad +
\left(\mathcal{H}^2_{RS}-\mathcal{H}_{RR}\mathcal{H}_{SS}\right)R \, ,\\
&\hat P=\mathcal{H}_{RR} \mathcal{H}_{SS}- \mathcal{H}^2_{RS}\,.
\end{split}
\end{equation}

In the following it will be convenient to use the three-vector notation
of (\ref{eq:EB}) and (\ref{eq:DH}).
%introduce a three-dimensional notation and to 
%split the electromagnetic field strength
%$F_{ab}$ as well as the excitation $H_{ab}$ into electric and magnetic parts, 
%$\boldsymbol{E},\boldsymbol{B}$ and $\boldsymbol{D},\boldsymbol{H}$. 
We decompose each of the 3-vectors $\boldsymbol{E}$, $\boldsymbol{B}$, $\boldsymbol{D}$,
and $\boldsymbol{H}$ into amplitude and direction, 
\begin{equation}\label{eq:vw}
\begin{split}
&\boldsymbol B (x^\mu,t) =B(x^\mu,t)\,\boldsymbol v(x^\mu,t)\, ,\\[0.1cm]
&\boldsymbol E (x^\mu,t) =E(x^\mu,t)\,\boldsymbol w(x^\mu,t)\, ,\\[0.1cm]
&\boldsymbol H (x^\mu,t) =H(x^\mu,t)\,\boldsymbol r(x^\mu,t)\, ,\\[0.1cm]
&\boldsymbol D (x^\mu,t) =D(x^\mu,t)\,\boldsymbol s(x^\mu,t) \, ,
\end{split}
\end{equation}
where $|\boldsymbol v|=|\boldsymbol w|=|\boldsymbol r|=|\boldsymbol s|=1$.
The spatial and temporal parts of $F^{im}F^k_{\phantom km}p_ip_k$, which enter into (\ref{eq:a1a2}),
can then be written as
\begin{gather}
\nonumber
F^{\alpha m}F^\beta_{\phantom \beta m}p_\alpha p_\beta = 
B^2\left[\boldsymbol p \cdot \boldsymbol p-\left(\boldsymbol v \cdot \boldsymbol p\right)^2\right]-E^2\left(\boldsymbol w \cdot \boldsymbol p\right)^2
\\
\label{eq:FF}
F^{4 m}F^\beta_{\phantom \beta m}p_\beta =  B\,E\, \boldsymbol{p} \cdot (\boldsymbol{w}\times\boldsymbol{v})
\\[0.1cm]
\nonumber
F^{4 m}F^4_{\phantom 4 m} = E^2 \, .
\end{gather}
Similarly, 
%as well as the parts of $\tilde H{}^{im}\tilde H{}^{j}_{\phantom jm}p_ip_j$:
\begin{gather}
\nonumber
\tilde H{}^{\alpha m}\tilde H{}^{\beta}_{\phantom \beta m}p_\alpha p_\beta= 
D^2\left[\boldsymbol p \cdot \boldsymbol p-\left(\boldsymbol s \cdot \boldsymbol p\right)^2\right]-H^2\left(\boldsymbol r \cdot \boldsymbol p\right)^2
\\
\label{eq:FFt}
\tilde H{}^{4 m}\tilde H{}^{\beta}_{\phantom \beta m}p_\beta =  D\,H\,\boldsymbol{p} \cdot (\boldsymbol{s}\times\boldsymbol{r})
\\[0.1cm]
\nonumber
\tilde H{}^{4 m}\tilde H{}^{4}_{\phantom 4 m} = H^2
\end{gather}
which will be used later.

%----------------------------------------------------------------------------------------------------------------------------------------
\subsection{Rayvelocity and Hamilton equations for the rays}
Interpreting $Q_A = a_A^{ik}p_ip_k$ as a Hamiltonian, the characteristic partial differential 
equation $a_A^{ik}\partial _i \psi \partial _k \psi =0$ can be viewed as a Hamilton-Jacobi 
equation. The corresponding set of Hamilton equations, or canonical equations, determines the 
\emph{bicharacteristic curves} or \emph{rays}. For background material on the notions of 
characteristics and bicharacteristics we refer to Courant and Hilbert \cite{CourantHilbert1962}.

The rays are defined with respect to $Q_1$ and $Q_2$ separately, i.e., they depend on the 
polarization. The canonical equations read
\begin{equation}
\frac{\mathrm dx^a}{\mathrm ds}=\frac{\partial Q_A}{\partial p_a}\, , \quad
\frac{\mathrm dp_a}{\mathrm ds}=-\frac{\partial Q_A}{\partial x^a} \, .
\end{equation}
Here $s$ is a parameter along the rays which has no obvious physical
meaning. In the following it will be convenient to reparametrize the rays by the time coordinate $t$,
cf.~\cite{CourantHilbert1962}. In order to do this, we have to assume that the rays of the Hamiltonian
$Q_A$ are causal (i.e., timelike or lightlike) with respect to the Minkowski background metric. It was 
shown by Obukhov and Rubilar~\cite{ObukhovRubilar2002} that the optical metrics are always of Lorentzian 
signature, provided that we exclude the pathological cases where they degenerate. However, no 
convenient criterion on the Lagrangian $\mathcal{L}(F,G)$ seems to be known that guarantees causality of the
rays with respect to the background metric. We will investigate this question in a separate paper;
here we just restrict our discussion, from now on, to Lagrangians where the rays of the optical metrics are 
causal with respect to the background Minkowski metric.
 
Then it is guaranteed that $a_A^{44} < 0$ and we may write the optical metrics as
%\footnote{That the light-cone of the optical metric does not enter the space-like area of the 
%Minkowski-metric is sufficient for a Lorentz-signature of the optical metric which is compatible 
%to Minkowsi's space-time-metric. Therefore it holds $a^{44}_A<0$.}
\begin{equation}\label{eq:fp}
a_A^{ik}p_ip_k = \frac{a_A^{44}}{c^2} (c\,p_4+H_A^+)(c\,p_4+H_A^-)
\end{equation}
where
\begin{equation}\label{eq:HA}
H_A^{\pm} = \, c \, \left( \frac{a_A^{\alpha 4}p_\alpha}{a_A^{44}}\pm
\sqrt{\left(\frac{a_A^{\alpha 4}p_\alpha}{a_A^{44}}\right)^2-
\frac{a_A^{\alpha\beta}p_\alpha p_\beta}{a_A^{44}}} \, \right)\, .
\end{equation}
Equation (\ref{eq:fp}) corresponds to splitting the null cone of the optical metric $a_A^{ik}$ into a future and 
a past cone. If we restrict our work here to future-oriented rays, we can write the characteristic equation as
\begin{equation}\label{eq:HAp}
c \, p_4 \, + \,  H_A^+ \, = \, 0
\end{equation} 
and the canonical equations read
\begin{equation}\label{eq:Halpha}
\frac{\mathrm dx^\alpha}{\mathrm dt}\, = \,  \frac{\partial H^+_A}{\partial p_\alpha} \, ,
\quad
\frac{\mathrm dp_\alpha}{\mathrm dt} \, = \, -  \, \frac{\partial H^+_A}{\partial x^\alpha} \, ,
 \end{equation}
\begin{equation}\label{eq:H4}
\frac{\mathrm dx^4}{\mathrm dt}=c \, , \quad
\frac{\mathrm dp_4}{\mathrm dt}=- \,\frac{\partial H_A^+}{\partial x^4} \, .
\end{equation}

\noindent
If $a_A^{ik}$ is known, integration of (\ref{eq:Halpha}) gives the spatial paths of the rays. 
The first equation of (\ref{eq:H4}) says that the new parameter $t$ coincides with the coordinate
time, while the second equation gives the change of the frequency of light.

The ray velocity can be read from (\ref{eq:Halpha}), 
\begin{equation}
v_{\mathrm{S}}^\alpha:=\frac{\mathrm dx^\alpha}{\mathrm dt}=\frac{\partial H^+_A}{\partial p_\alpha}\,.
\end{equation}

\noindent
The phase velocity (\ref{eq:vP}) can be rewritten in terms of the Hamiltonian as
\begin{equation}\label{eq:vp2}
v_{\mathrm{P}}^\beta =-\frac{c\,p_4}{{p_\alpha p^\alpha}}\,p^{\beta}=
\frac{H^+_A}{{p_\alpha p^\alpha}}\,p^{\beta}\,.
\end{equation}
Phase and ray velocity coincide if and only if 
\begin{equation}
\frac{H_A^{+}}{p_\alpha p^\alpha}\,p^\beta=\frac{\partial H_A^+}{\partial p_\beta}
\end{equation}
which is true if and only if $H^+_A$ is of the form
\begin{equation}\label{GeschBeding}
H_A^+ \, = \, f(x^\mu,ct) \, \sqrt{p_{\alpha} p^{\alpha}}
\end{equation}
where $f$ is any function of the space-time coordinates.
Equation (\ref{GeschBeding}) is satisfied in the usual vacuum theory of 
Maxwell but not in general in other $\mathcal{L}(F,G)$ theories. 
Note that (\ref{GeschBeding}) implies
\begin{equation}
\frac{\mathrm dx^\alpha}{\mathrm dt}=\frac{\partial H_A^+}{\partial p_\alpha}
=\frac{f(x^{\mu},ct) \, p^\alpha}{\sqrt{p^\beta p_\beta}}\,,
\end{equation}
i.e., the condition $v^\alpha_{\mathrm{S}}=v^\alpha_{\mathrm{P}}$ can hold only if $dx^{\alpha}/dt$ 
and $p^{\alpha}$ are parallel.

%------------------------------------------------------------------------------------------------------------
\subsection{Parallel electric and magnetic fields}\label{subsec:par}
We consider now the special case that $\boldsymbol{E}$ and $\boldsymbol{B}$ are 
parallel, i.e., that $\boldsymbol{v} = \boldsymbol{w}$ in the notation of (\ref{eq:vw}). 
This case covers, of course,  
in particular the situation that one of the two field strengths, $\boldsymbol{E}$ or 
$\boldsymbol{B}$, is zero. With the aid of the transformation (\ref{DualRotTab}), 
described in the Appendix, we will then discuss, at the end 
of this section, the case that the excitations $\boldsymbol{D}$ and $\boldsymbol{H}$
are parallel.

If we specialize (\ref{eq:FF}) to the case $\boldsymbol{v} = \boldsymbol{w}$ and insert the
result into (\ref{eq:aA}), the optical metrics read
\begin{equation}\label{eq:QApar}
\begin{split}
&a_A^{ik} p_i p_k = - (1- \sigma_A E^2) p_4^2 
+(1 + \sigma_A B^2) \boldsymbol{p}^2 
\\
&\qquad\qquad\qquad\qquad\quad\qquad
-\sigma _A (B^2+E^2)(\boldsymbol{w} \cdot \boldsymbol{p})^2 \, .
\end{split}
\end{equation}
Hence, the Hamiltonian $H_A^+$ from (\ref{eq:HA}) simplifies to
\begin{equation}\label{eq:HApar}
H_A^+ = c \, \sqrt{ \dfrac{(1+\sigma _A B^2)}{(1-\sigma _A E^2)} 
\Big(  | \boldsymbol{p} | ^2 - ( \boldsymbol{w} \cdot \boldsymbol{p})^2 \Big) + 
( \boldsymbol{w} \cdot \boldsymbol{p} )^2 \;} 
\end{equation}
and the phase velocity (\ref{eq:vp2}) reads
\begin{equation}\label{eq:vppar}
v_{\mathrm{P}} = c \, \sqrt{  \dfrac{(1+\sigma _A B^2)}{(1-\sigma _A E^2)}\,
\left( 1 - \dfrac{( \boldsymbol{w} \cdot \boldsymbol{p})^2}{ | \boldsymbol{p} |^2 }\right)
+  \dfrac{( \boldsymbol{w} \cdot \boldsymbol{p})^2}{ | \boldsymbol{p} |^2 }} \, .
\end{equation}
If we assume, in addition, that the unit vector of the background field is homogeneous, 
$\partial_\alpha w^\beta=0$, and that the amplitudes of the field strengths  
change only in the direction of $\boldsymbol p$, $\text{grad}\,B \propto \boldsymbol p$ 
as well as $\text{grad}\,E \propto \boldsymbol p$ , the canonical equations (\ref{eq:Halpha}) 
reduce to
\begin{equation}\label{eq:Halphapar}
\begin{split}
&\frac{\mathrm dx^\alpha}{\mathrm dt}=\frac{c}{H^+_A}  \left\{
\dfrac{(1+\sigma _A B^2)}{(1-\sigma _A E^2)} 
\Big( p^{\alpha} - ( \boldsymbol{w} \cdot \boldsymbol{p} )\,  
w^{\alpha} \Big) \right.
\\
&\qquad\qquad\qquad\quad+\left.( \boldsymbol{w} \cdot \boldsymbol{p} ) \, 
w^{\alpha}\right\} \,;\qquad\frac{\mathrm dp_\alpha}{\mathrm dt}\propto p_\alpha\,.
\end{split}
\end{equation}
The last equation implies that the direction of $p_\alpha$ is preserved along the ray. 

If additionally the background fields are static, $\partial \boldsymbol{E} / \partial t = 
\boldsymbol{0}$ and $\partial \boldsymbol{B} / \partial t = \boldsymbol{0}$,  the second
equation of (\ref{eq:H4}) reduces to 
\begin{equation}\label{eq:H4stat}
\frac{\mathrm dp_4}{\mathrm dt}=0 
%\quad\Leftrightarrow\quad \frac{\mathrm d\omega}{\mathrm dt}=0
\end{equation}
which means that, in this case, the background fields do not change the frequency of light. 

We are now interested in the special case that (\ref{GeschBeding}) holds which guarantees that 
phase velocity and ray velocity are equal and that $dx^{\alpha}/dt$ is parallel to $p^{\alpha}$. 
As the direction of $p^{\alpha}$ is preserved, the ray must then be a straight line.

There are two main cases where the Hamiltonian takes the form of 
(\ref{GeschBeding}). First, if $\boldsymbol{p} \, || \, \boldsymbol{w}$, 
we find from (\ref{eq:HApar}), (\ref{eq:vppar}) and (\ref{eq:Halphapar}) 
that $H_A^+=c \, | \boldsymbol{p}|$, $v_{\mathrm{P}} = c$, and 
$\mathrm{d}x^{\alpha}/ \mathrm{d} t = c \, p^{\alpha} / | \boldsymbol{p}|$, 
i.e., in this case the background fields have no effect. 
Second, if $\boldsymbol{p} \cdot \boldsymbol{w} =0$,
one gets
\begin{eqnarray}
&&\label{eq:vpmax1}H_A^+ \, = \, c \, | \boldsymbol{p} | \,
\sqrt{  \dfrac{1+\sigma _A B^2}{1-\sigma _A E^2}\,} \, ,
\\[0.1cm]
&&\label{eq:vpmax3} v_{\mathrm{P}} 
\, = \, c \, \sqrt{  \dfrac{1+\sigma _A B^2}{1-\sigma _A E^2}} \, , 
\\[0.1cm]
&&\label{eq:vpmax2}\frac{\mathrm dx^\alpha}{\mathrm dt}=
c \, \sqrt{  \dfrac{1+\sigma _A B^2}{1-\sigma _A E^2}}
\, \dfrac{p^{\alpha}}{|\boldsymbol{p}|} \, .
\end{eqnarray}
This is the case which is most appropriate for the proposed experiment, because in this case one 
achieves two goals: the rays do not deviate from a straight line but the phase 
velocity does change in comparison to the Maxwell standard vacuum theory. 

Now we go over to the case that $\boldsymbol{D}$ and $\boldsymbol{H}$ are parallel, i.e.., that
$\boldsymbol r=\boldsymbol s$. With the help of (\ref{DualRotTab}) from the Appendix 
we find that in this case (\ref{eq:HApar}) and  (\ref{eq:vppar}) 
have to be replaced with 
\begin{equation}
H_A^+ = c \, \sqrt{ 
\dfrac{(1+ \hat{\sigma}{}_A D^2)}{(1- \hat{\sigma}{}_A H^2)} 
\Big(  | \boldsymbol{p} | ^2 - ( \boldsymbol{s} \cdot \boldsymbol{p})^2 \Big) + 
( \boldsymbol{s} \cdot \boldsymbol{p} )^2 \;
} 
\end{equation}
\begin{equation}
v_{\mathrm{P}} = c \, \sqrt{  \dfrac{(1+\hat{\sigma} _A D^2)}{(1-\hat{\sigma} _A H^2)} 
\,\Big( 1 - \dfrac{( \boldsymbol{s} \cdot \boldsymbol{p})^2}{| \boldsymbol{p}|^2 } \Big)
+\dfrac{( \boldsymbol{s} \cdot \boldsymbol{p})^2}{| \boldsymbol{p}|^2 } } \, .
\end{equation}
As above one gets for homogeneous and time-independent excitations
\begin{equation}
\begin{split}
&\frac{\mathrm dx^\alpha}{\mathrm dt}=\frac{c}{H^+_A}  \left\{
\dfrac{(1+\hat{\sigma} _A D^2)}{(1-\hat{\sigma} _A H^2)} 
 \right.\\
&\qquad\qquad\qquad\quad\left.\times\Big( p^{\alpha} - (\boldsymbol{s} \cdot \boldsymbol{p} )  s^{\alpha} \Big) + 
( \boldsymbol{s} \cdot \boldsymbol{p} ) \, s^{\alpha} \, \right\} \, ,\\
&\frac{\mathrm dp_\alpha}{\mathrm dt} \propto  p_{\alpha} 
\, , \quad 
\frac{\mathrm dp_4}{\mathrm dt}=0 \,.
\end{split}
\end{equation}
Again, the case that $\boldsymbol p$ is parallel to $\boldsymbol s$ leads to $v_{\mathrm{P}}=c$, so this case is of no 
interest for us. If, however, $\boldsymbol p\cdot\boldsymbol s=0$, we get
\begin{eqnarray}
&&H_A^+ \, = \, c \, | \boldsymbol{p} | \,
\sqrt{  \dfrac{1+ \hat{\sigma}{}_A D^2}{1- \hat{\sigma}{}_A H^2}\,} \, ,
\\[0.1cm]
&&\label{eq:vpmax4} v_{\mathrm{P}} 
\, = \, c \, 
\sqrt{  \dfrac{1+ \hat{\sigma}{}_A D^2}{1- \hat{\sigma}{}_A H^2}} \, ,
\\[0.1cm]
&&\frac{\mathrm dx^\alpha}{\mathrm dt}=
c \, \sqrt{  \dfrac{1+ \hat{\sigma}{}_A D^2}{1- \hat{\sigma}{}_A H^2}}
\, \dfrac{p^{\alpha}}{|\boldsymbol{p}|} 
\end{eqnarray}
and there is no deviation of a light ray from a straight line.

%------------------------------------------------------------------------------------------------------------
\section{An interferometric experiment for testing nonlinear electrodynamics}\label{experiment}
There are two ways in which Michelson interferometry can be used for testing nonlinear
electrodynamics. First, a strong background field could be applied to the light beam in one arm of
the interferometer. One would compare the situation where the background field is switched on with 
the situation where it is switched off, cf.~\cite{Boer}. Second, one could place the whole 
interferometer in a strong background field. One would then search for changes in the interference 
pattern if the interferometer is being rotated. The first possibility is reasonable if one thinks of 
a large interferometer, with an arm
length of several meters at least. The second possibility is 
reasonable if one thinks of a tabletop interferometer. As an alternative to using a traditional Michelson
interferometer, one could also use a pair of optical resonators as they have been used for 
high-precision Michelson-Morley experiments in recent years. As these 
resonators have a typical size of only a few centimeters, one would do the 
experiment with the whole instrument placed in a background field. With the 
resonators oriented perpendicularly to each other, one would then compare 
the situation where the field is switched on with the situation where it is 
switched off, or one would rotate the whole instrument with keeping the 
field switched on.
       
In the following we first discuss the setup of the experiment where  a traditional Michelson
interferometer is used and the field is placed in one arm. This is the variant which 
brings out the basic idea of the experiment most clearly. Later in this section we discuss 
the other variants. 

\begin{figure}
\begin{center}
\includegraphics[width=9cm]{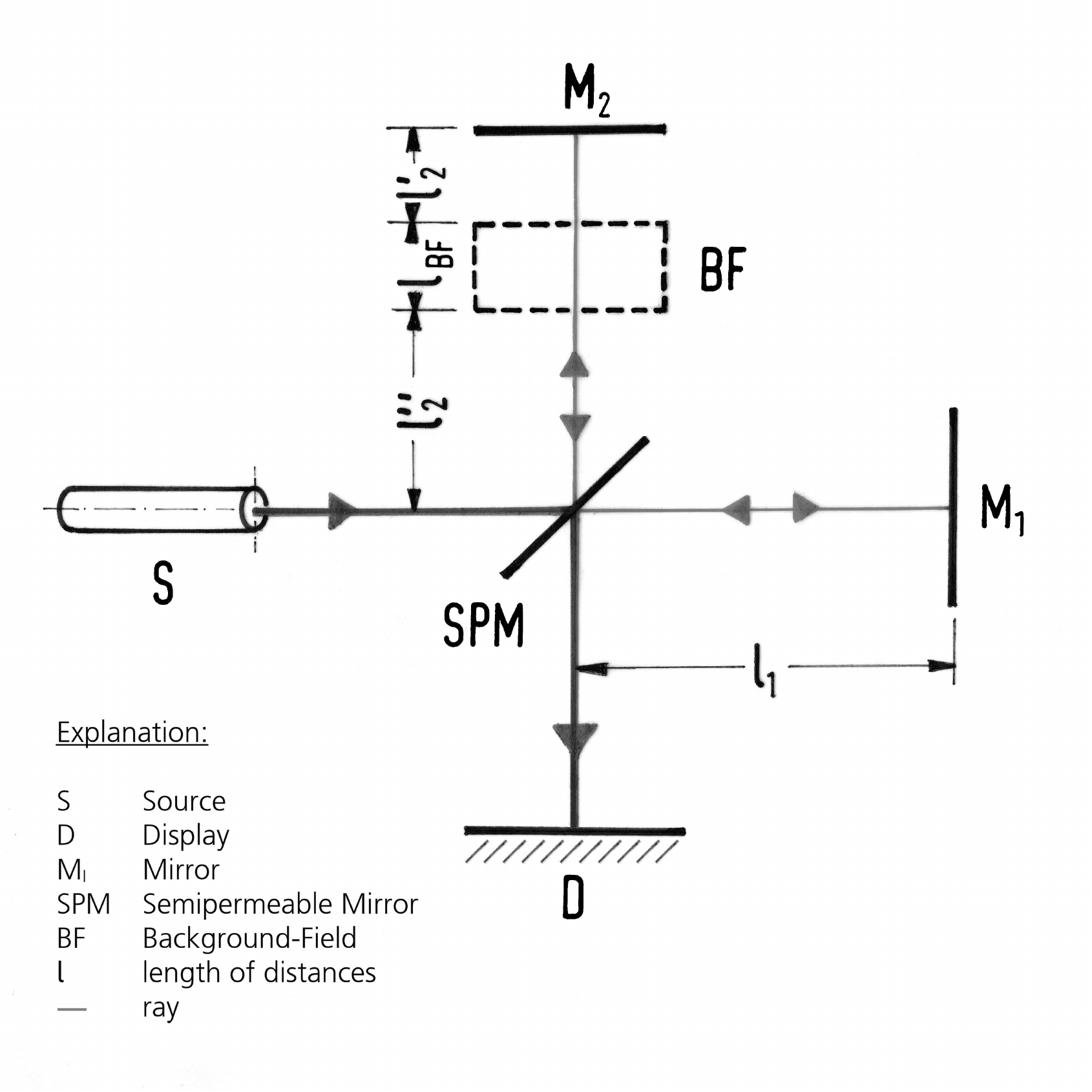}
\end{center}
\caption{Experimental setup}
\label{MichelsonBild}
\end{figure}

Figure~\ref{MichelsonBild} shows the interferometer with 
the background field in the region denoted $BF$. The ray leaves the source $S$ and is divided 
at the semipermeable mirror $SPM$. After reflection at the mirrors $M_1$ and $M_2$, respectively,
both parts interfere at $D$. If the background field is switched off, both parts always
travel with the standard vacuum phase velocity $c$. If the background field is switched on,
the part which travels along $l_2$ crosses the region $BF$ with a different phase velocity,
according to nonlinear electrodynamics. This would lead to a change of the interference 
pattern.

We consider the background field to be static, with one of the four fields $\boldsymbol{E}$,
$\boldsymbol{B}$, $\boldsymbol{D}$, or $\boldsymbol{H}$ vanishing. Each of these four cases
is covered by the calculations of the preceding section. We assume that the background field
is perpendicular to the propagation direction of the light. We have seen that in this
situation the ray does not deviate from a straight line. 
 
\vspace{0.2cm}

Obviously the travel times of the ray along the different sections are given by
\begin{equation}
ct_1=l_1\,,\quad ct'_2=l'_2\,,\quad 
ct''_2=l''_2\,,\quad 
v_{\mathrm{P}} \, t_{BF}=l_{BF}\,.
\end{equation}
Without background field the phase velocity is equal to $c$ everywhere, including the region $BF$.
The time delay $\Delta t_I$ of the two arms is therefore given by
\begin{equation}
\Delta t_I= 2(t_1-t'_2-t''_2-t_{BF})=\frac{2}{c}(l_1-l'_2-l''_2-l_{BF})\,.
\end{equation}
With background field the phase velocity in the region $BF$ is $v_{\mathrm{P}}$ which is, in general,
different from $c$.
The time delay $\Delta t_{II}$ of the two arms is therefore given by
\begin{equation}
\begin{split}
\Delta t_{II}&= 2(t_1-t'_2-t''_2-t_{BF})\\
&=\frac{2}{c}(l_1-l'_2-l''_2-\frac{c}{v_{\mathrm{P}}}l_{BF})\,.
\end{split}
\end{equation}
The change of the interference pattern is given by the time difference
\begin{equation}
\Delta t= \Delta t_{II}-\Delta t_I=\frac{2\,l_{BF}}{c}\left(1-\frac{c}{v_{\mathrm{P}}}\right)\,.
\end{equation}
This leads to a line shift of
\begin{equation}\label{eq:Delta}
\Delta= \frac{\omega \, \Delta t}{2 \, \pi}
=\frac{\omega\,l_{BF}}{\pi \, c}\left(1-\frac{c}{v_{\mathrm{P}}}\right)\,.
\end{equation}
Here $\omega$ denotes the frequency of the light. Note that $\omega$ is a constant because
the background field is assumed static.

% \footnote{The former considerations ensure that $\omega$ is constant along the whole 
% way through the interferometer, because a static background field ensures the validity 
% of $\partial B/\partial t=0$.}

%This phase shift leads to a line shift, measured in band size units, of
%\begin{equation}
%\Delta=\frac{\Delta\phi}{2\pi}=\frac{\omega\,l_{HF}}{c\,\pi}\left(1-\frac{c}{v_{\mathrm{P}}}\right)=\frac{2\,l_{HF}}{\lambda}\left(1-\frac{c}{v_{\mathrm{P}}}\right)\,.
%\end{equation}
%Here $\lambda$ is the wave-length of the light when leaving the source, i.e. $\lambda\omega=2\pi c$. 
%In the region $BF$ the wave-length is, of course, different.

We evaluate the general result for each of the four cases 
$\boldsymbol{E=0}$, $\boldsymbol{B=0}$, $\boldsymbol{D=0}$ 
and $\boldsymbol{H=0}$. Note that in general $\boldsymbol{E=0}$ 
is \emph{not} equivalent to $\boldsymbol{D=0}$ and $\boldsymbol{B=0}$ 
is \emph{not} equivalent to $\boldsymbol{H=0}$.

\begin{description}
\item[a) Magnetostatic field strength $\boldsymbol{(E=0)}$]
\hfill

\parindent=-0.4cm
From (\ref{eq:vpmax3}) we find that
\begin{equation}
\begin{split}
v_{\mathrm{P}} &=c\,\sqrt{1+\sigma_{1/2}B^2}\\[0.1cm]
&=c\left(1+\sigma_{1/2}(0)\frac{B^2}{2} + \, \dots \, \right)
\end{split}
\end{equation}

\parindent=-0.4cm
and hence, by (\ref{eq:Delta}),
\begin{equation}
\begin{split}
\Delta &=\frac{\omega\,l_{BF}}{\pi \, c}
\left(1-\frac{1}{\sqrt{1+\sigma_{1/2}B^2}}\right)
\\[0.1cm]
&=\frac{\omega \, l_{BF}\sigma_{1/2}(0)B^2}{2 \, \pi \, c}
+ \, \dots \, 
\end{split}
\end{equation}
\item[b) Electrostatic field strength $\boldsymbol{(B=0)}$]
\hfill

\parindent=-0.4cm
From (\ref{eq:vpmax3}) we find that
\begin{equation}
\begin{split}
v_{\mathrm{P}} &=c\,\frac{1}{\sqrt{1-\sigma_{1/2}E^2}}\\[0.1cm]
&=c\left(1+\sigma_{1/2}(0)\frac{E^2}{2} + \, \dots \, \right ) 
\end{split}
\end{equation}

\parindent=-0.4cm
and hence, by (\ref{eq:Delta}),
\begin{equation}
\begin{split}
\Delta&=\frac{\omega\,l_{BF}}{c\,\pi}\left(1-\sqrt{1-\sigma_{1/2}E^2}\right)\\[0.1cm]
&=\frac{\omega \, l_{BF}\sigma_{1/2}(0)E^2}{2 \, \pi \, c} + \, \dots 
\end{split}
\end{equation}
\item[c) Magnetostatic excitation $\boldsymbol{(D=0)}$]
\hfill

\parindent=-0.4cm
From (\ref{eq:vpmax4}) we find that
\begin{equation}
\begin{split}
v_{\mathrm{P}} &=c\,\frac{1}{\sqrt{1-\hat\sigma_{1/2}H^2}}\\[0.1cm]
&=c\left(1+\hat\sigma_{1/2}(0)\frac{H^2}{2} + \, \dots \, \right)
\end{split}
\end{equation}

\parindent=-0.4cm
and hence, by (\ref{eq:Delta}),
\begin{equation}
\begin{split}
\Delta&=\frac{\omega\,l_{BF}}{c\,\pi}\left(1-\sqrt{1-\hat\sigma_{1/2}H^2}\right)\\[0.1cm]
&=\frac{\omega \, l_{BF}\hat\sigma_{1/2}(0)H^2}{2 \, \pi \, c} 
+ \, \dots
\end{split}
\end{equation}
\item[d) Electrostatic excitation $\boldsymbol{(H=0)}$]
\hfill

\parindent=-0.4cm
From (\ref{eq:vpmax4}) we find that
\begin{equation}
\begin{split}
v_{\mathrm{P}} &=c\,\sqrt{1+\hat\sigma_{1/2}D^2}\\[0.1cm]
&=c\left(1+\hat\sigma_{1/2}(0)\frac{D^2}{2} + \, \dots \, \right)
\end{split}
\end{equation}

\parindent=-0.4cm
and hence, by (\ref{eq:Delta}),
\begin{equation}
\begin{split}
\Delta&=\frac{\omega\,l_{BF}}{c\,\pi}
\left(1-\frac{1}{\sqrt{1+\hat \sigma_{1/2}D^2}}\right)
\\[0.1cm]
&=\frac{\omega \, l_{BF}\hat\sigma_{1/2}(0)D^2}{2 \, \pi \, c}
+\, \dots 
\end{split}
\end{equation}
\end{description}

If one writes $X$ for $E$, $B$, $D$, or $H$, one can combine all results up to first order
in the form 
\begin{equation}\label{Allg1OrdPhasenG}
v_{\mathrm{P}}=c\left(1+\overset{X}\sigma_{1/2}(0)
\frac{X^2}{2} + \, \dots \, \right)
\end{equation}
and 
\begin{equation}\label{Allg1OrdDelta}
\Delta=\frac{\omega \, l_{BF}
\overset{X}\sigma_{1/2}(0)X^2}{2 \, \pi \, c}+ \, \dots 
\end{equation}
Here $\overset{X}\sigma_{1/2}(0)$ denotes either $\sigma_{1/2}(0)$ or $\hat\sigma_{1/2}(0)$,
depending on whether $X$ is a field strength or an excitation. Note that $2\pi c/\omega$
is the wavelength in the case of a vanishing background field. According to nonlinear 
electrodynamics, the wavelength changes when the ray travels through the background field $BF$. 
This means that, if one substitutes the angular frequency $\omega$ by the wavelength $\lambda=2\pi c/\omega$,
\begin{equation}\label{Allg1OrdDelta_lambda}
\Delta=\frac{l_{BF}\overset{X}\sigma_{1/2}(0)X^2}{\lambda}+ \, \dots \, \,,
\end{equation} 
one has to keep in mind that $\lambda$ is not the wavelength of the light when passing through 
the background field but of the light when emitted by the source.

The results of this section can also be applied to the case where the whole interferometer is
inside the background field. Here one does not switch on and off the background field but 
rotates the interferometer by $90\degree$ so that in the initial position the first
arm is orthogonal to the field and the second arm is parallel to the field while in
the end position it is vice versa. Then one gets instead of the preceding formulas the 
following ones:
\begin{equation}
\Delta t=\frac{2\left(l_1+l_2\right)}{c}\left(1-\frac{c}{v_p}\right) \, ,
\end{equation}
\begin{equation}
\Delta=\frac{\omega \, \left(l_1+l_2\right)
\overset{X}\sigma_{1/2}(0)X^2}{2 \, \pi \, c}+ \, \dots 
\end{equation}
This means that one has to replace $l_{BF}$ by $l_1+l_2$ in all formulas to go 
from the fist setup to the second one. 

As an alternative to using a traditional Michelson interferometer
with two arms, we will now discuss a setup
with optical resonators as it has been used frequently
in recent years for high-precision Michelson interferometry;
see e.g. \cite{SH} and the references therein. Here one
uses a laser which is stabilized to the eigenfrequency
$\nu _{\mathrm{eigen}}  = N v_{\mathrm{P}}/(2L)$ of an 
optical resonator, where $N$ is the mode number, 
$v_{\mathrm{P}}$ is the phase velocity of light and $L$ 
is the length of the resonator. 
The quality of a resonator is determined by its finesse $F$,
typically $F = 100 \, 000$. In a figurative way, a resonator 
may be viewed as equivalent to a traditional interferometer
whose arm length is folded $F$ times.  

If $v_{\mathrm{P}}$ and $L$ 
undergo a change, the eigenfrequency of the resonator and
therefore the frequency of the stabilized laser changes as
\begin{equation}
\frac{\delta\nu}{\nu}=\frac{\delta v_{\mathrm{P}}}{v_{\mathrm{P}}}
-\frac{\delta L}{L}\,.
\end{equation}
If the resonator is put into a homogeneous and static $\boldsymbol{E}$,
$\boldsymbol{B}$, $\boldsymbol{D}$, or $\boldsymbol{H}$ field, with 
its axis perpendicular to the field, the phase 
velocity of light changes according to 
\begin{equation}
\frac{\delta v_{\mathrm{P}}}{v_{\mathrm{P}}}
\approx \overset{X}\sigma_{1/2}(0)\frac{X^2}{2}\,.
\end{equation}
if we use the approximations of (\ref{Allg1OrdPhasenG}). As a direct 
measurement of $\delta \nu$ is not possible, one superimposes to the
first laser a second reference laser stabilized to the eigenfrequency
$\nu _{\mathrm{ref}}$ of a resonator with (ideally) the same physical 
characteristics as the first one. Then the difference of the frequencies 
$\Delta \nu := \nu _{\mathrm{eigen}} - \nu _{\mathrm{ref}}$ appears as 
the carrier frequency of the resulting beat. If the second resonator 
is oriented parallel to the background field and thus not influenced 
by it, this means that $\Delta \nu = \delta \nu$.

For a theoretical discussion of the effect, we assume
that $L$ is not changed if the background field is applied.
(Of course, for a practical realization of the experiment
one has to take into account that the material of the
resonator is influenced, e.g., by magnetostriction, but
we ignore this here.) Then
\begin{equation}
\frac{\delta \nu}{\nu}=
\frac{\delta v_{\mathrm{P}}}{v_{\mathrm{P}}}
\approx \overset{X}\sigma_{1/2}(0)\frac{X^2}{2}\,.
\end{equation}
In  \cite{SH}, by averaging over many measurements it was 
possible  to determine $\delta \nu / \nu$ 
with an accuracy of $10^{-17}$. If we assume that
the same accuracy can be reached in the experiment 
proposed here, a measurable effect requires that X satisfies
\begin{equation}\label{eq:nuacc}
\frac{\delta v_{\mathrm{P}}}{v_{\mathrm{P}}}
\approx 
\overset{X}\sigma_{1/2}(0)\frac{X^2}{2}
\approx 10^{-17} \, .
\end{equation}

In the next section we discuss the perspectives of performing 
such an experiment as a test of particular theories of the
Pleba{\'n}ski class. We compare the setup with the background
field placed in one arm of a big Michelson interferometer
with the setup using optical resonators. In the following, we
refer to the first one as the "large-scale experiment" 
and to the second one as the "small-scale experiment."

%-------------------------------------------------------------------------------------------------------------------
\section{Application to special theories of the Pleba{\' n}ski class}\label{konkretes}

\subsection{Born-Infeld theory}\label{subsec:BI}
In the case of the Born-Infeld theory, Lagrangian and Hamiltonian are given 
by~\cite{BornInfeld1934}
\begin{equation}\label{BILagrangeFkt1}
\mathcal{L}(F,G)=-b_0^2\sqrt{1+\frac{F}{b_0^2}-\frac{G^2}{b_0^4}}+b_0^2 \, ,
\end{equation}
\begin{equation}\label{BILagrangeFkt2}
\mathcal{H}(R,S)=b_0^2\sqrt{1+\frac{R}{b_0^2}-\frac{S^2}{b_0^4}}-b_0^2 \, ,
\end{equation}
where $b_0$ is a new constant of Nature with the dimension of a 
field strength. The constitutive law reads
\begin{equation}\label{BIFeldstaerke1}
H^{ab}=\frac{F^{ab}- \frac{G}{b_0^2} \tilde F{}^{ab}}{\sqrt{1+\frac{F}{b_0^2}-\frac{G^2}{b_0^4}}}
\end{equation}
which can be solved for the field strength,
\begin{equation}\label{BIFeldstaerke2}
F^{mn}=\frac{H^{mn}+\frac{S}{b_0^2}\,\tilde H{}^{mn}}{\sqrt{1+\frac{R}{b_0^2}-\frac{S^2}{b_0^4}}}\,.
\end{equation}
The invariants $R$ and $S$ are given in terms of $F$ and $G$ by
\begin{gather}
\frac{R}{b^2_0}=\frac{-\frac{F}{b_0^2}+4\frac{G^2}{b_0^4}
+\frac{F}{b_0^2}\frac{G^2}{b_0^4}}{1+\frac{F}{b_0^2}-\frac{G^2}{b_0^4}} \, ,\\[0.1cm]
S=-G \, ,
\end{gather}
which implies
\begin{equation}
\frac{1+\frac{F}{b_0^2}-\frac{G^2}{b_0^4}}{1+\frac{G^2}{b_0^4}}
=\frac{1+\frac{S^2}{b_0^4}}{1+\frac{R}{b_0^2}-\frac{S^2}{b_0^4}}\, .
\end{equation}
This leads to
\begin{eqnarray}
\nonumber&&\mathcal{L}_{FF}=-\frac{2\,\mathcal{L}_F^3G}{b_0^2}\,,
\quad \mathcal{L}_{FG}=\frac{4\,\mathcal{L}_F^3G}{b_0^4}\,,\\
&&\mathcal{L}_{GG}=-\frac{2\,\mathcal{L}_F}{b_0^2}-\frac{8\,\mathcal{L}_F^3G^2}{b_0^2}\, ,\\
\nonumber
\quad&&\mathcal{H}_{RR}=-\frac{2\,\mathcal{H}_R{}^3S}{b_0^2}\,,
\quad \mathcal{H}_{RS}=\frac{4\,\mathcal{H}_R{}^3S}{b_0^4}\,,\\
&&\mathcal{H}_{SS}=-\frac{2\,\mathcal{H}_R}{b_0^2}-\frac{8\,\mathcal{H}_R{}^3 S^2}{b_0^2}\, .
\end{eqnarray}
Therefore one gets for the functions $\sigma_{1/2}$ and $\hat\sigma_{1/2}$, 
which give the deviation from the standard Maxwell vacuum theory,
\begin{equation}\label{eq:sigmaBI}
\begin{split}
&\sigma_1=\sigma_2=-\frac{1}{b_0^2+F}=-\frac{1}{b_0^2}+\cdots\\
\quad&\hat \sigma_1=\hat \sigma_2=-\frac{1}{b_0^2+R}=-\frac{1}{b_0^2}+\, \dots 
\end{split}
\end{equation}

It is worth noticing that $\sigma_1=\sigma_2$ holds not only in the Born-Infeld
theory and in the standard vacuum Maxwell theory but also in any other theory whose 
Lagrangian differs only by a term linear in G from them. (Such theories, however,
are often excluded because they are not invariant under parity transformations.)

Additionally one can calculate the phase velocity and the line shift for the four 
static cases:
\begin{description}
\item[Cases a) with $\mathbf{E=0}$ and d) with $\mathbf{H=0}$]
\quad\\
Here we discuss the case of a magnetostatic field strength and the case of an electrostatic 
excitation together. If we use the abbreviation $Y=B,D$, we find 
\begin{equation}
\overset{Y}\sigma_1=\overset{Y}\sigma_2=\frac{-1}{b_0^2+Y^2} \, ,
\end{equation}
hence
\begin{equation}
v_{\mathrm{S}}=v_{\mathrm{P}}= \dfrac{c}{\sqrt{1+\frac{Y^2}{b_0^2}}}
=
c\left(1-\frac{Y^2}{2b_0^2}+\cdots\right) . \!\!\!\!\!\!\!
\end{equation}
The limit $Y\rightarrow0$ yields $v_{\mathrm{S}}=v_{\mathrm{P}}\rightarrow c$ as it has to.

By contrast, the limit $Y\rightarrow \infty$ yields $v_{\mathrm{S}}=v_{\mathrm{P}}\rightarrow 0$, so one may 
say that the background field $Y$ slows down the light ray. There is no upper bound
for $Y=B,D$. 

The line shift is given by
\begin{equation}
\begin{split}
\Delta&=\frac{2l_{BF}}{\lambda}\left(1-\sqrt{1+\frac{Y^2}{b_0^2}} \right)\\
&=-\frac{l_{BF}}{\lambda}\frac{Y^2}{b_0^2}+\, \dots 
\end{split}
\end{equation}

\item[Cases b) with $\mathbf{B=0}$ and c) with $\mathbf{D=0}$]
\quad\\
Here we discuss the case of an electrostatic field strength and the case of a magnetostatic excitation together. 
If we use the abbreviation $Z=E,H$, we find  
\begin{equation}
\overset{Z}\sigma_1=\overset{Z}\sigma_2=\frac{1}{Z^2-b_0^2} \, ,
\end{equation}
hence
\begin{equation}
v_{\mathrm{S}}=v_{\mathrm{P}}= c \, \sqrt{1 - \frac{Z^2}{b_0^2}}
=c\left(1-\frac{Z^2}{2b_0^2}+\cdots\right). \!\!\!\!\!\!\!\!\!\!
\end{equation}
Again, the limit $Z\rightarrow0$ yields $v_{\mathrm{S}}=v_{\mathrm{P}} \rightarrow c$. 

In contrast to the case above this one leads to an upper bound for $Z$. This is obvious 
because for $Z\rightarrow b_0$ one gets $v_{\mathrm{S}}=v_{\mathrm{P}} \rightarrow 0$. In analogy to the other 
cases a background excitation slows down the light ray.
For background fields $Z>b_0$ one gets an imaginary phase velocity, so one has to conclude that
$Z\leq b_0\,.$

The line shift is given by
\begin{equation}
\begin{split}
\Delta&=\frac{2l_{BF}}{\lambda}
\left(1- \dfrac{1}{\sqrt{1 - \frac{Z^2}{b_0^2}}} \right) \\
&=-\frac{l_{BF}}{\lambda}\frac{Z^2}{b_0^2}+\, \dots 
\end{split}
\end{equation}
\end{description}

We may combine the first-order approximations of all preceding 
cases into two formulas, one for the velocity and one for the line shift:
\begin{equation}
v_{\mathrm{S}}=v_{\mathrm{P}} \approx c\left(1-\frac{X^2}{2b_0^2}\right)\quad\text{and}\quad\Delta\approx-\frac{l_{BF}}{\lambda}\frac{X^2}{b_0^2}\,.
\end{equation}
To give an example we calculate the line shift for the large-scale experiment for some specific values. We assume an 
accuracy of about $10^{-6}$ line shifts. For $l_{BF}=100\,\mathrm{m}$ and 
$\lambda = 1000\,\mathrm{nm}$ one gets:
\begin{equation}
\Delta=-\frac{l_{BF}}{\lambda}\frac{X^2}{b_0^2}\approx-10^8 \,\frac{X^2}{b_0^2}\,.
\end{equation}
With these values one sees an effect if 
\begin{equation}\label{eq:Xest}
X \gtrsim 10^{-7} \, b_0 \, .
\end{equation}
Born and Infeld conjectured that 
\begin{equation}
b_0=\frac{e}{r_e^2} \approx 6 \times 10^{15}\,\frac{\sqrt{\mathrm g}}{\sqrt{\mathrm{cm}}\,\mathrm s}
\end{equation}
where $e$ is the electron charge and $r_e$ is the classical electron radius. Although
this is only of historical interest, we remark that the corresponding line shift would be
\begin{equation}
\Delta\approx - \frac{10^{-22}}{36}\,X^2\,\frac{\mathrm{cm}\,\mathrm{s}^2}{\mathrm{g}}\,.
\end{equation}
If this were true we would need a field strength or an excitation of 
\begin{equation}
X \gtrsim 6 \times 10^{8}\,\frac{\sqrt{\mathrm{g}}}{\sqrt{\mathrm{cm}}\,\mathrm{s}}
\end{equation}
to see an effect. For a magnetic field strength, $X = B$, this would correspond to $B_{\mathrm{SI}} \gtrsim
6 \times 10^4 T$ in SI units, see (\ref{eq:SI}). Clearly, this is not achievable in the foreseeable future.

It is more interesting to see what lower bound on $b_0$ one could get from an
experiment. Let us assume that
\begin{equation}\label{eq:Xest2}
X\approx \times 10^{4}\,\frac{\sqrt{\mathrm{g}}}{\sqrt{\mathrm{cm}}\,\mathrm{s}}
\end{equation}
which corresponds to $B_{\mathrm{SI}}= 1\, \mathrm{T}$ according to (\ref{eq:SI}).
This is not an unrealistic value for a magnetic field to be produced in a laboratory.
% along a distance of $100\,\mathrm{m}$. For example this can be done with a Halbach 
% cylinder (compare Grote~\cite{Grote}). 
Then a null result of our large-scale experiment would imply, according to (\ref{eq:Xest}), that 
\begin{equation}
b_0 \gtrsim 1 \times 10^{10}\,\frac{\sqrt{\mathrm{g}}}{\sqrt{\mathrm{cm}}\,\mathrm{s}}\,.
\end{equation}
For the small-scale experiment we suggest a lower field strength of $300\,\mathrm{mT}$ to 
prevent magnetorestriction. Then we find 
from (\ref{eq:nuacc}) and (\ref{eq:sigmaBI}) that
\begin{equation}
b_0 \gtrsim 7 \times 10^{11} 
\frac{\sqrt{\mathrm{g}}}{\sqrt{\mathrm{cm}}\,\mathrm{s}}\,
\end{equation}
which is almost 2 orders of magnitude better than the large-scale 
experiment.

%----------------------------------------------------------------------------------------
\subsection{Born's theory}
Born's  Langrangian~\cite{Born1933} differs from the Born-Infeld Lagrangian by omitting the 
$G^2$ term,
\begin{equation}
\mathcal{L}(F)=-b_0^2\sqrt{1+\frac{F}{b_0^2}}+b_0^2\,.
\end{equation}
This leads to an electrodynamical theory with birefringence,
\begin{eqnarray}
\sigma_1=0 \, , \qquad
\sigma_2=\frac{-1}{b_0^2+F}\,.
\end{eqnarray}
From the viewpoint of geometrical optics Born's theory is a hybrid.
One polarization mode behaves according to the standard vacuum Maxwell theory
and the other one according to the Born-Infeld theory.
This means that, if one filters the $\sigma_2$ rays out with a polarization
filter, then one sees no difference to Maxwell, and if one filters 
the $\sigma_1$ rays out, then one sees no difference to Born-Infeld. 
As  a consequence, the results of Sec.~\ref{subsec:BI} are also valid 
for the $\sigma_2$ rays in Born's theory.

%----------------------------------------------------------------------------------------
\subsection{Series expansions for electrodynamics with arbitrary Lagrangian}\label{ArbLag}
If we are interested only in first-order deviations from Maxwell's theory, 
we may express the Langrangian in terms of a series expansion with respect
to $F/A$ and $G/A$ up to second order, where $A$ is a constant with the 
dimension of a field strength squared. Introducing $A$ is necessary because
only for dimensionless terms is it meaningful to say that they are small
without referring to a particular system of units. In the Born-Infeld theory, 
e.g., we choose $A=b_0^2$. 

The series expansion of the Lagrangian reads 
\begin{equation}\label{AllemLagFkt}
\begin{split}
&\mathcal{L}=\alpha+\underset{1}\beta\,\frac{F}{A}+\underset{2}\beta\,\frac{G}{A}+\\
&\qquad\qquad+\underset{1}\gamma\left(\frac{F}{A}\right)^2
+\underset{2}\gamma \frac{F\,G}{A^2}+\underset{3}\gamma \left(\frac{G}{A}\right)^2+\, \dots 
\end{split}
\end{equation}
Note that $\underset{2}\beta$ and $\underset{2}\gamma$ are zero if
the theory is invariant under parity transformations. One can assume 
the validity of the following bookkeeping system for the smallness of 
terms, where $\sim$ means that terms are of the same order.
\begin{itemize}
\item $\alpha\sim\underset{i}\beta\sim\underset{i}\gamma\cdots$ as well as 
$F^{mn} \sim \tilde{F}{}^{mn}$, hence $F\sim G$.
\item $F/A$ and $G/A$ are dimensionless with $F/A\sim G/A$.
\item From the first order of $R=R(F,G)$ and $S=S(F,G)$ [cf. 
(\ref{HamiltonBez_1a}) to (\ref{HamiltonBez_2d})] one gets 
$\underset{i}\beta F/A\sim AR/\underset{i}\beta$.
\end{itemize}

With the help of (\ref{HamiltonBez_1a}) to (\ref{HamiltonBez_2d})) one can 
now calculate $R$ and $S$ as series in $F$ and $G$. Additionally one can then calculate 
the inverted series, i.e. $F$ and $G$ as a series in $R$ and $S$. This step allows us then 
to calculate the Hamiltonian as a function of $R$ and $S$. The result of this calculation is
\begin{equation}
\begin{split}
&\mathcal{H}=-\alpha+\underset{1}B\,\frac{R}{\hat A}+\underset{2}B\,\frac{S}{\hat A}+
\\
&\qquad\qquad+\underset{1}C\,
\left(\frac{R}{\hat A}\right)^2
+\underset{2}C\,\frac{R\,S}{\hat A^2}
+\underset{3}C\,\left(\frac{S}{\hat A}\right)^2+\cdots\,,
\end{split}
\end{equation}
with the following coefficients:
\begin{eqnarray}
\nonumber&&\hat A^{-1}:=\frac{A}{\underset{2}{\beta }^2+4 \underset{1}{\beta }^2}\,;\quad\underset{1}B:=-\underset{1}{\beta }\,;\quad\underset{2}B:=\underset{2}{\beta }\,;\\
\nonumber&&\underset{1}C:=\left(-\underset{2}{\beta }^4 \underset{1}{\gamma }+2 \underset{1}{\beta } \underset{2}{\beta }^3 \underset{2}{\gamma }-4 \underset{1}{\beta }^2 \underset{2}{\beta }^2 \underset{3}{\gamma }+8 \underset{1}{\beta }^2 \underset{2}{\beta }^2 \underset{1}{\gamma }-\right.\\
\nonumber&&\qquad\qquad\left.-8 \underset{1}{\beta }^3 \underset{2}{\beta } \underset{2}{\gamma }-16 \underset{1}{\beta }^4 \underset{1}{\gamma }\right)\Big/\left(\underset{2}{\beta }^2+4 \underset{1}{\beta }^2 \right)^2\,;\\
&&\underset{2}C:=\left(-\underset{2}{\beta }^4 \underset{2}{\gamma }+4 \underset{1}{\beta } \underset{2}{\beta }^3 \underset{3}{\gamma }-16 \underset{1}{\beta } \underset{2}{\beta }^3 \underset{1}{\gamma }+24 \underset{1}{\beta }^2 \underset{2}{\beta }^2 \underset{2}{\gamma }-\right.\\
\nonumber&&\qquad\qquad\left.-16 \underset{1}{\beta }^3 \underset{2}{\beta } \underset{3}{\gamma }+64 \underset{1}{\beta }^3 \underset{2}{\beta } \underset{1}{\gamma }-16  \underset{1}{\beta }^4 \underset{2}{\gamma }\right)\Big/\left( \underset{2}{\beta }^2+4 \underset{1}{\beta }^2 \right)^2\,;\\
\nonumber&&\underset{3}C:=\left(-\underset{2}{\beta }^4 \underset{3}{\gamma }-8 \underset{1}{\beta } \underset{2}{\beta }^3 \underset{2}{\gamma }+8 \underset{1}{\beta }^2 \underset{2}{\beta }^2 \underset{3}{\gamma }-64 \underset{1}{\beta }^2 \underset{2}{\beta }^2 \underset{1}{\gamma }+\right.\\
\nonumber&&\qquad\qquad\left.+32 \underset{1}{\beta }^3 \underset{2}{\beta }  \underset{2}{\gamma }-16 \underset{1}{\beta }^4 \underset{3}{\gamma }\right)\Big/\left( \underset{2}{\beta }^2+4 \underset{1}{\beta }^2 \right)^2\,.
\end{eqnarray}
This leads to some additions to the bookkeeping system:
\begin{itemize}
\item From $\underset{i}\beta F/A\sim AR/\underset{i}\beta$ one gets $F/A\sim R/\hat A$ as well as $G/A\sim S/\hat A$.
\item It is easy to see that $\alpha\sim\underset{i}\beta\sim\underset{i}\gamma\cdots\sim\underset{i}B\sim\underset{i}C\cdots$ holds.
\end{itemize}
As a result of this bookkeeping system one sees that, if one 
can neglect in the Lagrangian terms of a certain
order in $F/A$ and $G/A$, then one can neglect in the Hamiltonian 
terms of the same order in $R/\hat{A}$ and $S/\hat{A}$. 

For the case of a parity-invariant Lagrangian, $\mathcal{L}(F,G)=\mathcal{L}(F,-G)$, 
the coefficients of the Hamiltonian become very simple:
\begin{equation}
\begin{split}
&\underset{1}B=-\underset{1}\beta\,;\quad \underset{2}B=-\underset{2}\beta=0\,;\quad\\
&\underset{1}C=-\underset{1}\gamma\,;\quad\underset{2}C=-\underset{2}\gamma=0\,;\quad \underset{3}C=-\underset{3}\gamma\,,\\
&\hat A=\frac{4\underset{1}\beta^2}{A}\,.
\end{split}
\end{equation}
If, in addition, the first-order approximation of the Lagrangian coincides
with the standard Maxwell one (\ref{eq:LagMax}) --- which is true if $\underset{1}\beta=- A/2$ ---
one gets
\begin{equation}
\hat A=A\,.
\end{equation}
This is, in particular, the case for the theories of Born, Born-Infeld, and 
Heisenberg-Euler. 

From the Lagrangian (or the Hamiltonian, respectively) one gets the 
``deviation coefficients" $\sigma_{1/2}$ in the zeroth order of approximation 
with respect to $F$ and $G$ (or $R$ and $S$, respectively):
\begin{eqnarray}
\nonumber\sigma_{1/2}(0)\,&&=\frac{2 \underset{1}{\gamma }}{A \underset{1}{\beta }}+\frac{ \underset{3}{\gamma }}{2 A \underset{1}{\beta }}\\
\label{sigma1OrdFl}&&\qquad\pm\frac{1}{2} \sqrt{\frac{16  \underset{1}{\gamma }^2+4  \underset{2}{\gamma }^2-8  \underset{1}{\gamma } \underset{3}{\gamma }+ \underset{3}{\gamma }^2}{A^2 \underset{1}{\beta }^2}}\\
\nonumber\hat\sigma_{1/2}(0)\,&&=\frac{2 \underset{1}{C}}{\hat A \underset{1}{B}}+\frac{\underset{3}{C}}{2 \hat A\underset{1}{B}}\\
\label{sigma1OrdEr}&&\qquad\pm\frac{1}{2} \sqrt{\frac{16  \underset{1}{C}^2+4  \underset{2}{C}^2-8  \underset{1}{C} \underset{3}{C}+ \underset{3}{C}^2}{\hat A{}^2  \underset{1}{B}^2}}\,.
\end{eqnarray}

Obviously the zeroth order approximation of $\sigma_{1/2}$ gives the first-order 
approximation of the optical metric for the deviation from Maxwell's theory and 
therefore also of the phase velocity.

The Born-Infeld theory, e.g.,  yields $\sigma_1(0)=\sigma_2(0)=\hat\sigma_1(0)=
\hat\sigma_2(0)=-1/b_0^2$, so one recovers the values calculated above. Additionally 
one sees that the approximation procedure does not destroy the absence of birefrigence 
in the given order of approximation.

One gets the results for the four cases described in Sec.~\ref{experiment} if one 
feeds (\ref{sigma1OrdEr}) and (\ref{sigma1OrdFl}) into 
(\ref{Allg1OrdPhasenG}) and (\ref{Allg1OrdDelta}):

\begin{description}
\item[Magnetostatic field strength case ($E=0$)]
\begin{equation}
\begin{split}
v_{\mathrm{P}} =c&+c\,\frac{B^2}{2}\left(\frac{2 \underset{1}{\gamma }}{A \underset{1}{\beta }}+\frac{ \underset{3}{\gamma }}{2 A \underset{1}{\beta }}\right.\\
&\qquad\left.\pm\frac{1}{2} \sqrt{\frac{16  \underset{1}{\gamma }^2+4  \underset{2}{\gamma }^2-8  \underset{1}{\gamma } \underset{3}{\gamma }+ \underset{3}{\gamma }^2}{A^2 \underset{1}{\beta }^2}}\right)+\, \dots 
\end{split}
\end{equation}
\begin{equation}
\begin{split}
\Delta=&\frac{l_{BF}\,B^2}{\lambda}\left(\frac{2 \underset{1}{\gamma }}{A \underset{1}{\beta }}+\frac{ \underset{3}{\gamma }}{2 A \underset{1}{\beta }}\right.\\
&\qquad\quad\left.\pm\frac{1}{2} \sqrt{\frac{16  \underset{1}{\gamma }^2+4  \underset{2}{\gamma }^2-8  \underset{1}{\gamma } \underset{3}{\gamma }+ \underset{3}{\gamma }^2}{A^2 \underset{1}{\beta }^2}}\right)+\, \dots 
\end{split}
\end{equation}

\item[Electrostatic field strength case ($B=0$)]
\begin{equation}
\begin{split}
v_{\mathrm{P}}=c&+c\,\frac{E^2}{2}\left(\frac{2 \underset{1}{\gamma }}{A \underset{1}{\beta }}+\frac{ \underset{3}{\gamma }}{2 A \underset{1}{\beta }}\right.\\
&\qquad\left.\pm\frac{1}{2} \sqrt{\frac{16  \underset{1}{\gamma }^2+4  \underset{2}{\gamma }^2-8  \underset{1}{\gamma } \underset{3}{\gamma }+ \underset{3}{\gamma }^2}{A^2 \underset{1}{\beta }^2}}\right)+\, \dots 
\end{split}
\end{equation}
\begin{equation}
\begin{split}
\Delta=&\frac{l_{BF}\,E^2}{\lambda}\left(\frac{2 \underset{1}{\gamma }}{A \underset{1}{\beta }}+\frac{ \underset{3}{\gamma }}{2 A \underset{1}{\beta }}\right.\\
&\qquad\quad\left.\pm\frac{1}{2} \sqrt{\frac{16  \underset{1}{\gamma }^2+4  \underset{2}{\gamma }^2-8  \underset{1}{\gamma } \underset{3}{\gamma }+ \underset{3}{\gamma }^2}{A^2 \underset{1}{\beta }^2}}\right)+\, \dots 
\end{split}
\end{equation}

\item[Magnetostatic excitation case ($D=0$)]
\begin{equation}
\begin{split}
v_{\mathrm{P}} =c&+c\,\frac{H^2}{2}\left(\frac{2 \underset{1}{C}}{\hat A \underset{1}{B}}+\frac{\underset{3}{C}}{2 \hat A\underset{1}{B}}\right.\\
&\,\,\left.\pm\frac{1}{2} \sqrt{\frac{16  \underset{1}{C}^2+4  \underset{2}{C}^2-8  \underset{1}{C} \underset{3}{C}+ \underset{3}{C}^2}{\hat A{}^2  \underset{1}{B}^2}}\right)+\, \dots 
\end{split}
\end{equation}
\begin{equation}
\begin{split}
\Delta=&\frac{l_{BF}\,H^2}{\lambda}\left(\frac{2 \underset{1}{C}}{\hat A \underset{1}{B}}+\frac{\underset{3}{C}}{2 \hat A\underset{1}{B}}\right.\\
&\quad\left.\pm\frac{1}{2} \sqrt{\frac{16  \underset{1}{C}^2+4  \underset{2}{C}^2-8  \underset{1}{C} \underset{3}{C}+ \underset{3}{C}^2}{\hat A{}^2  \underset{1}{B}^2}}\right)+\, \dots 
\end{split}
\end{equation}

\item[Electrostatic excitation case ($H=0$)]
\begin{equation}
\begin{split}
v_{\mathrm{P}} =c&+c\,\frac{D^2}{2}\left(\frac{2 \underset{1}{C}}{\hat A \underset{1}{B}}+\frac{\underset{3}{C}}{2 \hat A\underset{1}{B}}\right.\\
&\,\,\left.\pm\frac{1}{2} \sqrt{\frac{16  \underset{1}{C}^2+4  \underset{2}{C}^2-8  \underset{1}{C} \underset{3}{C}+ \underset{3}{C}^2}{\hat A{}^2  \underset{1}{B}^2}}\right)+\, \dots 
\end{split}
\end{equation}
\begin{equation}
\begin{split}
\Delta=&\frac{l_{BF}\,B^2}{\lambda}\left(\frac{2 \underset{1}{C}}{\hat A \underset{1}{B}}+\frac{\underset{3}{C}}{2 \hat A\underset{1}{B}}\right.\\
&\quad\left.\pm\frac{1}{2} \sqrt{\frac{16  \underset{1}{C}^2+4  \underset{2}{C}^2-8  \underset{1}{C} \underset{3}{C}+ \underset{3}{C}^2}{\hat A{}^2  \underset{1}{B}^2}}\right)+\, \dots 
\end{split}
\end{equation}
\end{description}

In principle it is easy to obtain further orders of approximation, but the
resulting terms are expected to be very small and will not be worked out here.

%--------------------------------------------------------------------------------------------------
\subsection{The Heisenberg-Euler theory}
Here we give an example for the procedure described in the preceding section. 
For small values of the field strength the Heisenberg-Euler theory can be described 
by the following Lagrangian~\cite{GD,HeisenbergEuler1936} which results from a series 
expansion with respect to $F$ and $G$:
\begin{equation}\label{eq:HEL}
\begin{split}
&\mathcal{L}=E_0^2\left\{-\frac12\,\frac{F}{E_0^2}+
\Lambda\left(\frac{F^2}{E^4_0}+7\,\frac{G^2}{E^4_0}\right)\right\}
\end{split}
\end{equation}
where
\begin{gather}
\Lambda=\frac{\hbar c}{90\pi e^2} = 0.7363 
\\
E_0=\frac{m^2 c^4}{e^3} = 
6.048 \times 10^{15}\,\frac{\sqrt{\mathrm g}}{\sqrt{\mathrm{cm}}\,\mathrm s} \, .
\end{gather}
Here $e$ is the electron charge, $m$ is the electron mass, $c$ is 
the speed of light and $\hbar$ is Planck's constant.

So the coefficients in (\ref{AllemLagFkt}) are
\begin{equation}
\begin{split}
&\alpha=0\,,\quad\underset{1}\beta=-\underset{1}B=-\frac{E_0^2}{2}\,,\quad\underset{2}\beta=0\,,\quad\\
&\underset{1}\gamma=-\underset{1}C=\Lambda\,E_0^2\,,\quad\underset{2}\gamma=-\underset{2}C=0\,,\quad\\
&\underset{3}\gamma=-\underset{3}C=7\Lambda\,E_0^2\,.
\end{split}
\end{equation}
and
\begin{equation}
\hat A=A=E_0^2\, .
\end{equation}
Hence
\begin{equation}
\begin{split}
&\sigma_1(0)=\hat\sigma_1(0)=-\frac{14\Lambda}{E_0^2}\,,\\
&\sigma_2(0)=\hat\sigma_2(0)=-\frac{8\Lambda}{E_0^2} \, .
\end{split}
\end{equation}
For the four possibilities for the background field described in 
Sec.~\ref{experiment} we get, using again the abbreviation $X=E,D,B,H$,
\begin{equation}
\begin{split}
v_{\mathrm{P}} (\sigma_1)&=c\left(1-\frac{7\Lambda\,X^2}{E_0^2}\right)+\cdots\\
v_{\mathrm{P}} (\sigma_2)&=c\left(1-\frac{4\Lambda\,X^2}{E_0^2}\right)+\cdots\,\\
\Delta(\sigma_1)&=-\frac{14 \,\Lambda\,l_{BF}X^2}{\lambda\,E_0^2}+\cdots\\
\Delta(\sigma_2)&=-\frac{8 \,\Lambda\,l_{BF}X^2}{\lambda\,E_0^2}+\, \dots 
\end{split}
\end{equation}

Using the same setup as before for the large-scale experiment, with 
$l_{BF}=100\,\mathrm{m}$ and $\lambda = 1000 \,\mathrm{nm}$ one gets
\begin{equation}
\begin{split}
\Delta(\sigma_1)\approx - 2 \times 10^{-23}\,X^2\,\frac{\mathrm{cm}\,\mathrm{s}^2}{\mathrm{g}}\,,
\\
\Delta(\sigma_2)\approx - 1\times 10^{-23}\,X^2\,\frac{\mathrm{cm}\,\mathrm{s}^2}{\mathrm{g}}\,.
\end{split}
\end{equation}
Therefore one needs a field strength or an excitation of
\begin{equation}
X \gtrsim 3 \times 10^{8}\,
\frac{\sqrt{\mathrm g}}{\sqrt{\mathrm{cm}}\,\mathrm s}
\end{equation}
to see any effect. This is clearly not achievable with present or near-future instruments.

A similar calculation shows that for the small-scale
experiment a field about 2 orders of magnitude
smaller would be sufficient. However, even in this case 
one would need a field of more than $10^{6}\,
\frac{\sqrt{\mathrm g}}{\sqrt{\mathrm{cm}}\,\mathrm s}\hat=100\,\mathrm{T}$
to see an effect.

%-----------------------------------------------------------------------------------------------
\section{Conclusions and Discussion}

Since Born and Infeld created their ``new field theory''
of electromagnetism \cite{BornInfeld1934}, different nonlinear 
modifications of vacuum electrodynamics on the basis of a
Lagrangian $\mathcal{L} (F,G)$ have been discussed, where usually
one considers only those theories that reproduce the standard
vacuum Maxwell theory in sufficiently weak fields.
All these new electrodynamical theories have in common
that they predict that light travels along the null cones of
two optical metrics, one for each polarization state, where
at least one of them differs from the vacuum Maxwell
light-cone. At the same time they introduce at least one
new dimensionfull constant of Nature.

While in the standard vacuum Maxwell theory the superposition
principle holds, this is no longer true in other
$\mathcal{L}(F,G)$ theories. As a consequence, an electromagnetic
background field would have an effect on the propagation
of electromagnetic waves and thus, in particular, on the
phase velocity of light. This is reflected by the fact that
the optical metrics depend on the background field. The
best technique for measuring small changes in the phase
velocity of light with high accuracy is interferometry. In
this paper we worked out the mathematical details for
using interferometry as a test of $\mathcal{L}(F,G)$ theories.

In cases where the constants of Nature that enter into
the theory are known, as e.g. in the Heisenberg-Euler
theory, an interferometric experiment could be used for
confirming the theory by verifying the prediction. If instead 
the constants of Nature that enter into
the theory are not known, as e.g. in the Born-Infeld theory,
a null result of the experiment would give bounds on
these constants. Our estimates demonstrate that, with
realistic (magnetic) fields, an interferometric experiment
could place significant bounds on the Born-Infeld constant
$b_0$.

Unfortunately, in the case of the Heisenberg-Euler theory
our estimates seem to indicate that a confirmation of the 
theory is not realizable with electromagnetic fields that 
can be achieved in present-day experiments. However, it might
be possible to considerably enhance the sensitivity by
using time-dependent background fields, rather than the 
static fields we have considered for our numerical estimates. 
For the case of 
testing the Heisenberg-Euler theory with an interferometer
of the size of a gravitational wave detector, this possibility
was discussed in detail recently by Grote \cite{Grote}.
The idea is to change the background field periodically with a
frequency $\omega$, e.g. by rotating a permanent magnet. 
As long as $\omega$ is small in comparison to the frequency
of the laser light used in the interferometer, our equations
could still be used for this situation in the sense of an 
adiabatic approximation. If the laser light is polarized, 
rotating the background field would lead to a 
periodically varying signal according to any theory that
predicts birefringence \textit{in vacuo}. (Unfortunately, this excludes
the Born-Infeld theory.)  By choosing long integration times
--- Grote suggests to run the experiment for a year ---
one could improve the statistics in such a way that 
it might be possible to reach the sensitivity for testing
the Heisenberg-Euler theory. A similar analysis has not
been carried through for the small-scale experiment so far.
We will leave this for other authors, as it goes beyond the
scope of the present paper which was to lay the theoretical 
foundations of the experiment in the context of an arbitrary 
$\mathcal{L} (F,G)$ theory.

Finally, we add a remark on pulsed background fields. 
Pulsed magnetic fields and also laser pulses (pulsed
null fields) can be produced with considerably higher
field strengths than static or slowly varying fields. 
For example, pulsed magnetic fields of $\approx 100\,\mathrm{T}$
have already been produced in the laboratory. However,
these fields persist only for short times, so the 
adiabatic approximation would not be valid which makes 
the theory considerably more difficult. Moreover, there
are several technical obstacles. For example, we see major experimental difficulties towards a realization of the small-scale experiment with (pulsed) magnetic fields of $\approx 100 \, \text{T}$
because of magnetostriction. Also, for the experiment
with a pulsed null field as a background one would wish
to have the pulse traveling in the same direction as the 
laser beam in the interferometer, to make sure that the 
latter does not deviate from a straight line. This cannot
be done without changing the geometry of the interferometer,
neither for the small-scale nor for the large-scale experiment. 
For these reasons, we have restricted our specific
calculations to time-independent background fields (which
includes the case of slowly varying fields in the sense
of an adiabatic approximation).

\section*{Acknowledgments}
G.S. wishes to thank Evangelisches Studienwerk Villigst for supporting him
with a Ph.D. stipend during the course of this work. V.P. is grateful to Deut\-sche
Forschungsgemeinschaft for financial support under Grant No. LA 905/14-1. Moreover, we gratefully 
acknowledge support from the Deutsche Forschungsgemeinschaft within the 
Research Training Group 1620 "Models of Gravity." We also thank Sven 
Herrmann for helpful discussions on the experimental aspects of the subject
and an anonymous referee for directing our attention to some
important references.
 
%---------------------------------------------------------------------------------------------------
\section*{Appendix: Hamiltonian formalism in terms of the excitation}
First we give a necessary and sufficient condition 
for the constitutive law (\ref{eq:Hab}) to be
locally solvable for $F^{ab}$. By the implicit function 
theorem, this is true if the Jacobian of the map from the 
field strength 6-vector to the excitation 6-vector is nonzero
After dividing by the factor 
$\left(4 \mathcal{L}_F^2 +\mathcal{L}_G^2\right){}^2$, which 
is nonzero unless the Lagrangian is constant and thus trivial,
we find that this condition reads
\begin{gather}\label{eq:Jac}
4 \mathcal{L}_F^2 \! +\mathcal{L}_G^2-
4 \left(F^2 \! +4 G^2\right) \mathcal{L}_{FF} \mathcal{L}_{GG}
+4 \left(F^2 \! +4 G^2\right) \mathcal{L}_{FG}^2
\nonumber
\\
+8 F \mathcal{L}_F \mathcal{L}_{FF} 
+16 G \mathcal{L}_F \mathcal{L}_{FG}
+4 F \mathcal{L}_{GG} \mathcal{L}_G 
-2 F \mathcal{L}_F \mathcal{L}_{GG}
\nonumber
\\
-8 G \mathcal{L}_{FF} \mathcal{L}_G 
+2 G \mathcal{L}_G \mathcal{L}_{GG} \neq0 \, .
\end{gather}
It is easy to see that this condition is satisfied, for all field 
configurations, in the Born theory and also in the Born-Infeld theory.
For the Heisenberg-Euler Lagrangian (\ref{eq:HEL}) it is true as well, 
where we have to observe that this second-order theory is valid only 
as long as the magnitude of the field strength is small in comparison 
to $E_0$.  

Whenever the constitutive law (\ref{eq:const}) can be solved for
$F_{mn}$, we can pass to a Hamiltonian description by a Legendre 
transformation (\ref{eq:Hamilton}). In this appendix we derive some 
relevant equations of the Hamiltonian formalism that will be used
in the body of the paper, based on an analogue 
formalism that was developed already by Born and 
Infeld~\cite{BornInfeld1934} for their special theory.

From (\ref{eq:Hamilton}) and (\ref{eq:const}) we find
\begin{equation}\label{eq:constH}
\dfrac{\partial \mathcal{H}}{\partial H^{ij}} =
- F_{ij} 
\end{equation}
which is the Hamiltonian version of the constitutive law.
In the case of vanishing sources, $j^m = 0$, the Maxwell equations
read
\begin{equation}\label{eq:Lvac}
\partial_nH^{mn}=0\quad\text{and}\quad\partial_{[a} F_{bc]}=0 \, .
\end{equation}
These two equations can be equivalently rewritten as
\begin{equation}\label{eq:Hvac}
\partial_{[a}\tilde H_{bc]}=0\quad\text{and}\quad\partial_n\tilde F{}^{mn}=0 \, .
\end{equation}
Comparison of (\ref{eq:const}) and (\ref{eq:Lvac}) on one side and
(\ref{eq:constH}) and (\ref{eq:Hvac}) on the other side demonstrates that
the source-free theory is invariant under a duality rotation
\begin{equation}\label{eq:duality}
F^{mn} \hookrightarrow \tilde H{}^{mn} \, , \quad
\mathcal{L} \hookrightarrow \mathcal{H} \, .
\end{equation}
In 3-vector notation, $F^{mn} \hookrightarrow \tilde H{}^{mn}$ means
$E_\alpha\hookrightarrow H_\alpha$ and $B_\alpha\hookrightarrow -D_\alpha$.
Clearly, $F^{mn} \hookrightarrow \tilde H{}^{mn}$ implies
\begin{equation}\label{DualRotTab}
\begin{split}
\tilde F{}^{mn}\hookrightarrow -H{}^{mn} \, , \quad
F\hookrightarrow R \, , \quad G\hookrightarrow S\,.
\end{split}
\end{equation}
If we start from the Lagrangian $\mathcal{L}(F_{mn})$ and
work out all relevant equations of the theory in terms of the 
field strength, we get the relevant equations in terms of the 
excitation simply by applying the replacements (\ref{eq:duality})
and (\ref{DualRotTab}). Note that this method works only in the 
case of vanishing sources, $j^m=0$, but for any Lagrangian
$\mathcal{L} (F_{mn})$ for which the constitutive law
(\ref{eq:const}) can be solved for $F_{mn}$.

We now specify to a Lagrangian of the Pleba{\'n}ski class. We
recall that in this case the constitutive law reads
\begin{equation}\label{HamiltonBez_1a}
H^{ab}=-2\,\mathcal{L}_F\,F^{ab}+\mathcal{L}_G\,\tilde F^{ab}\, .
\end{equation}
Similarly, (\ref{eq:constH}) specifies to
\begin{equation}\label{HamiltonBez_1b}
F_{ab}=2\,\mathcal{H}_R\,H_{ab}-\mathcal{H}_S\,\tilde H_{ab} \, .
\end{equation}
Inserting (\ref{HamiltonBez_1a}) into (\ref{eq:Hamilton}) yields
\begin{equation}\label{HamiltonBez_2a}
\mathcal{H}(R,S)=2\mathcal{L}_F F
+2 \mathcal{L}_G G-\mathcal{L}(F,G) \, 
\end{equation}
while inserting (\ref{HamiltonBez_1b}) into (\ref{eq:Hamilton}) yields
\begin{equation}\label{HamiltonBez_2b}
\mathcal{H}(R,S)=2 \mathcal{H}_R R
+2 \mathcal{H}_SS- \mathcal{L}(F,G) \, . 
\end{equation}
From these two equations we read that
\begin{equation}\label{HamiltonBez_1c}
\mathcal{L}_FF+\mathcal{L}_GG=\mathcal{H}_RR+\mathcal{H}_SS \, .
\end{equation}
Also, from (\ref{HamiltonBez_1a}) we find immediately that
\begin{equation}\label{HamiltonBez_2c}
\begin{split}
R=\left(-4\mathcal{L}_F^2+\mathcal{L}_G^2\right)^2F
-8\mathcal{L}_F\mathcal{L}_GG
\, , 
\\
S=\left(-4\mathcal{L}_F^2+\mathcal{L}_G^2\right)^2G
+2\mathcal{L}_F\mathcal{L}_GF \, .
\end{split}
\end{equation}
Similarly, from (\ref{HamiltonBez_1b}) we find that
\begin{equation}\label{HamiltonBez_2d}
\begin{split}
F=\left(-4\mathcal{H}_R^2+\mathcal{H}_S^2\right)^2R-
8\mathcal{H}_R\mathcal{H}_SS\,\, , 
\\
G=\left(-4\mathcal{H}_R^2+\mathcal{H}_S^2\right)^2S
+2\mathcal{H}_R\mathcal{H}_SR\,.
\end{split}
\end{equation}
In Sec.~\ref{ArbLag} the equations (\ref{HamiltonBez_1a}) to
(\ref{HamiltonBez_2d}) are used for calculating series expansions
of the Lagrangian and the Hamiltonian theory up to second order in 
$F$ and $G$. This enables one to calculate the first post-Maxwellian 
results of the discussed experiment for an arbitrary Lagrangian of 
the Pleba{\'n}ski class.

\end{document}